\documentclass{article}

\usepackage{arxiv}

\usepackage{amsmath,amsfonts}
\usepackage{algorithmic}
\usepackage{array}
\usepackage[caption=false,font=normalsize,labelfont=sf,textfont=sf]{subfig}
\usepackage{textcomp}
\usepackage{stfloats}
\usepackage{url}
\usepackage{hyperref}
\hypersetup{
    colorlinks=true,
    linkcolor=blue,
    citecolor=blue,
    filecolor=magenta,      
    urlcolor=cyan,
    pdftitle={Overleaf Example},
    pdfpagemode=FullScreen,
    }
\usepackage{verbatim}
\usepackage{graphicx}
\usepackage{siunitx}
\usepackage{float}
\usepackage{tabularx}

\def\BibTeX{{\rm B\kern-.05em{\sc i\kern-.025em b}\kern-.08em
    T\kern-.1667em\lower.7ex\hbox{E}\kern-.125emX}}
\usepackage{balance}
\begin{document}
\title{Low-dispersion low free-spectral-range Mach-Zehnder interferometer with long straight path lengths on silicon}
\author{\href{https://orcid.org/0000-0003-3493-527X}{\includegraphics[scale=0.06]{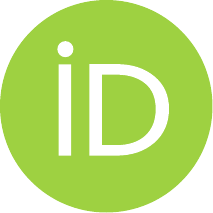}\hspace{1mm}Cael~Warner}\thanks{Electrical and Computer Engineering\\
University of Alberta\\
116 St \& 85 Ave, Edmonton T6G 2R3 \\
\texttt{spencerw@ualberta.ca} 
}}

\renewcommand{\shorttitle}{Low FSR MZI on Silicon}

\hypersetup{
pdftitle={Low dispersion low FSR MZI long path lengths Si},
pdfsubject={},
pdfauthor={Cael~Warner},
pdfkeywords={Mach-Zehnder, interferometer, silicon},
}

\maketitle

\begin{abstract}
Multiple Mach-Zehnder interferometers are constructed using fiber-Bragg grating couplers, y-branches, silicon waveguides, and/or broadband splitters in silicon on insulator strip waveguides to test the effect of variation in waveguide length between consecutive bends on the transmission gain spectrum, free-spectral range, group refractive index, and dispersion. Dispersion is mitigated by increasing the path length between consecutive bends of minimum radius. With long linear waveguide sections, a free spectral range of 0.41 \si{nm} is achieved.
\end{abstract}


\section{Introduction}
In a Mach-Zehnder interferometer (MZI), the path length difference between two parallel waveguides results in constructive and destructive interference which modifies their transmission spectrum with respect to wavelength. In this work, an MZI with a free spectral range (FSR) of only 0.41 \si{nm} shares excellent agreement with a simulated spectrum in ANSYS Optics INTERCONNECT, for a bandwidth of 10 \si{nm} from the center wavelength $\lambda=1550 \si{nm}$. Additional analytical fits yield excellent agreement with a correlation coefficient of $\mathrm{R}^2=0.91591$ across the full spectrum ranging from $\lambda=1500$ \si{nm} to $\lambda=1600$ \si{nm}, where $\lambda$ is the wavelength measured by the detector. 

Shorter path length MZIs transmission spectra are simulated and fitted to measured MZI spectra to measure group index $n_g$, effective index $n_{\mathrm{eff}}$, and dispersion, $D$. ANSYS Optics INTERCONNECT simulated spectra demonstrate a wavelength shift and variation of $n_{\mathrm{eff}}$ uncorrelated to path length, resulting in disagreement between measured and simulated transmission spectra, despite baseline subtraction of a de-embedding structure transmission spectrum. Therefore, dispersion is associated with the waveguide configuration. MZIs with greater path lengths between consecutive bends exhibit less dispersion and better agreement with the simulated spectra. Therefore, longer linear waveguides between minimum bend radii are preferable to mitigate dispersion and process variation.

\section{Narrow free spectral range Mach-Zehnder interferometer design}

\begin{figure}[H]
    \centering
    \includegraphics[width=0.5\linewidth]{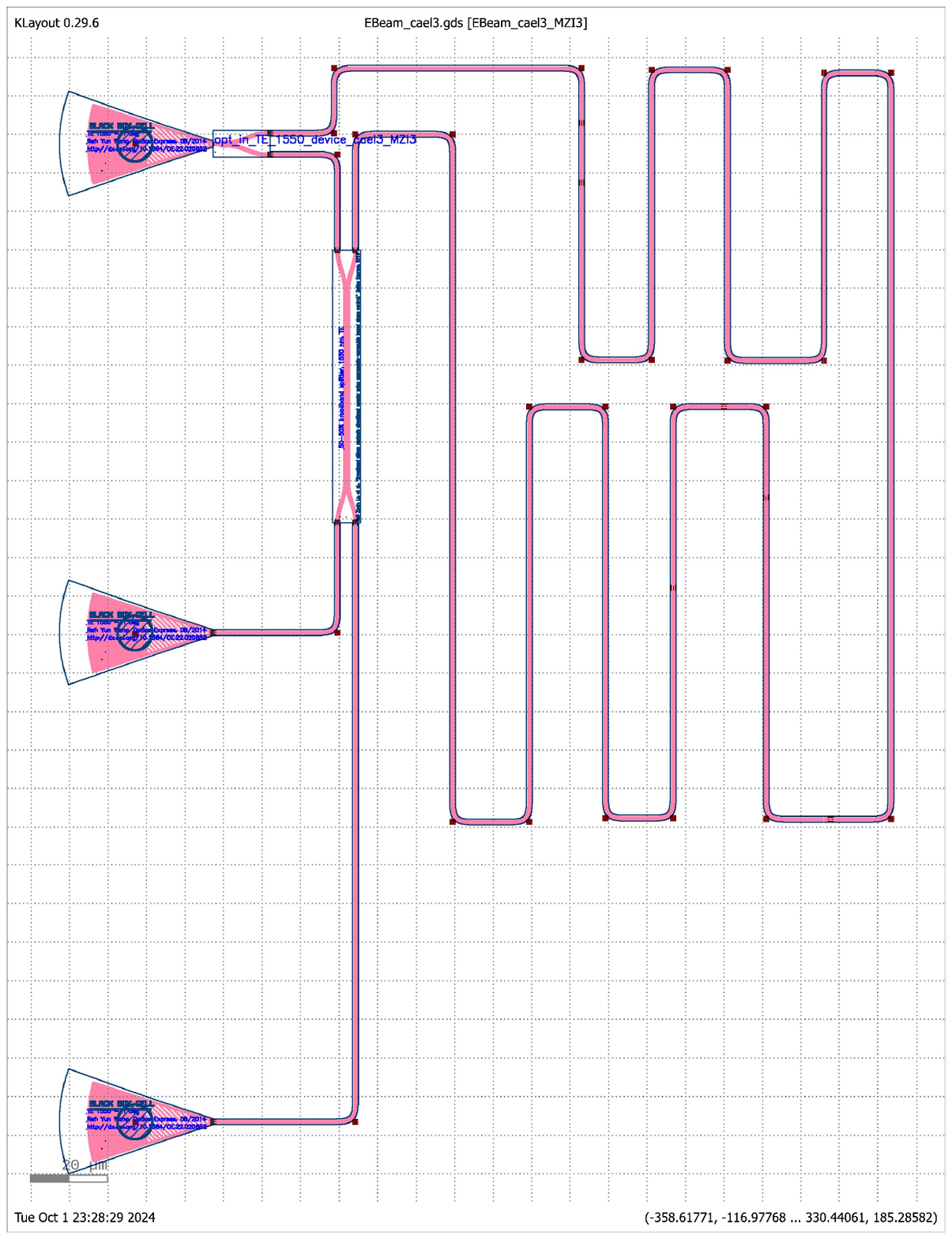}
    \caption{Narrow FSR MZI. Section $L_2$ consists of horizontal and vertical waveguide sections connected by standard 5 \si{\micro m} bends. The purpose of this design is to recover a narrow FSR of only 0.41 \si{nm} and examine its agreement with the simulated spectrum, and the corresponding bandwidth. }
    \label{fig:2}
\end{figure}

The MZI of interest has a narrow FSR of 0.41 \si{nm} and a floorplan illustrated in Fig. \ref{fig:2}. It uses a strip waveguide with standard 500 \si{nm} width, 220 \si{nm} height, and supports a single mode TE polarized wave. Fibre grating couplers are aligned to the left, with path length difference $\Delta L=L_2-L_1=1373.880$ \si{nm}.  Additional design iterations include a change in the path length between consecutive bends with the same minimum bend radius.

\begin{table}
    \centering
    \caption{Mach-Zehnder interferometer lengths}
    \begin{tabularx}{\textwidth}{|X|c|}
    \hline
         Waveguide lenth/MZI No. &(narrow FSR)\\ \hline
         $L_1$, \si{\micro m}& 40.788\\ \hline
         $L_2$, \si{\micro m}& 1414.668 \\ \hline
         $\Delta L$, \si{\micro m}&1373.880\\ \hline
    \end{tabularx}
    
    \label{tab:1}
\end{table}

\section{Variation in path length for Mach-Zehnder interferometer design}
Multiple MZIs, each with variable path length between consecutive bends, are fitted to determine a relation between $n_\mathrm{eff}$ and path length. Each bend radius is 5 \si{\micro m}, whereas the second branch length varies with respect to the first branch length. De-embedding structures help realize the insertion loss for appropriate fitting to each spectrum. Each MZI uses the same process as the narrow FSR MZI. Each MZI imbalance length $\Delta L=L_2-L_1$ is listed in Table \ref{tab:2}. FSR varies between 1.65 \si{nm} and 5.66 \si{nm}, as listed in Table \ref{tab:3}.

\begin{figure}
    \centering
    \includegraphics[width=0.75\linewidth]{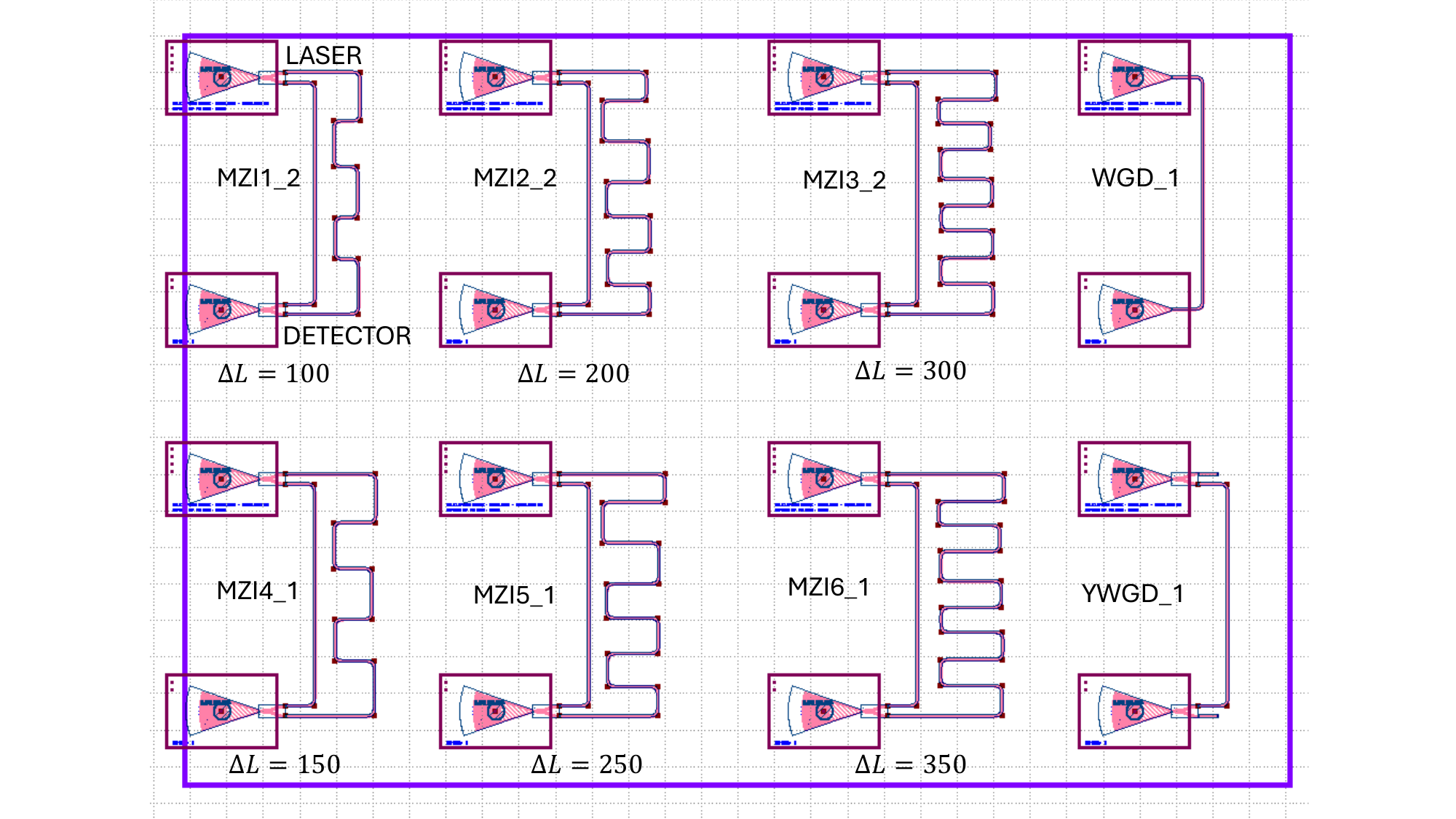}
    \caption{Varied path length Mach-Zehnder interferometer designs. Each of the MZIs have a path length difference on the order of 100-350 \si{\micro m}, a fraction the path length difference for the MZI shown in Fig. \ref{fig:2}. WGD and YWGD are de-embedding structures that include a path length of $L_1=150$ \si{\micro m} to represent the nominal path length that does not vary between the MZIs. }
    \label{fig:3}
\end{figure}

\begin{table}
    \centering
    \caption{Varying Mach-Zehnder interferometer path length difference}
    \begin{tabularx}{\textwidth}{|X|c|c|c|c|c|c|}
    \hline
         Waveguide lenth/MZI No. & 1  & 2 & 3 & 4 & 5 & 6\\ \hline
         $L_1$, \si{\micro m}& 150 & 150  & 150 & 150 & 150 & 150 \\ \hline
         $L_2$, \si{\micro m}& 250 & 350 & 450 & 300 & 400 & 500 \\ \hline
         $\Delta L$, \si{\micro m}& 100 & 200 & 300 & 150 & 250 & 350\\ \hline
    \end{tabularx}
    
    \label{tab:2}
\end{table}

\section{Fabrication}
The MZI structures are part of a NanoSOI multi-process wafer fabrication by Applied Nanotools Inc. (\href{http://www.appliednt.com/nanosoi}{http://www.appliednt.com/nanosoi}; Edmonton, Canada) based on direct-write 100 \si{keV} electron beam lithography technology. More details of fabrication are listed in the Supplementary material.

\section{Measurement description}
To characterize the MZIs, an automated test setup \cite{ref2,ref6} with automated control software written in Python was used \cite{ref3}.  An Agilent 81600B tunable laser was used as the input source and Agilent 81635A optical power sensors as the output detectors. The wavelength was swept from 1500 to 1600 \si{nm} in 10 \si{pm} steps.  A polarization maintaining (PM) fibre was used to maintain the polarization state of the light, to couple the TE polarization into the grating couplers \cite{ref4}.  A 90$^\circ$ rotation was used to inject light into the TM grating couplers \cite{ref4}.  A polarization maintaining fibre array was used to couple light in/out of the chip \cite{ref5}.

\section{Mach-Zehnder interferometer response}
\begin{figure}[H]
    \centering
    (a)\includegraphics[width=0.4\linewidth]{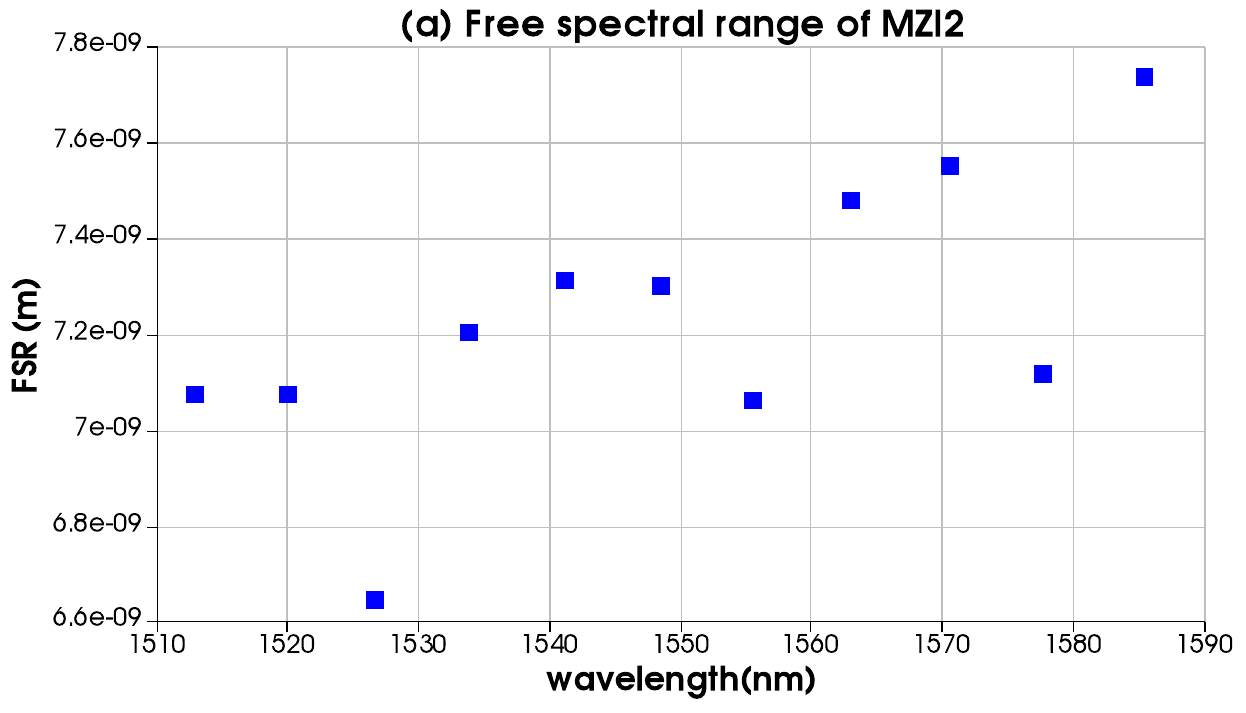}
    (b)\includegraphics[width=0.4\linewidth]{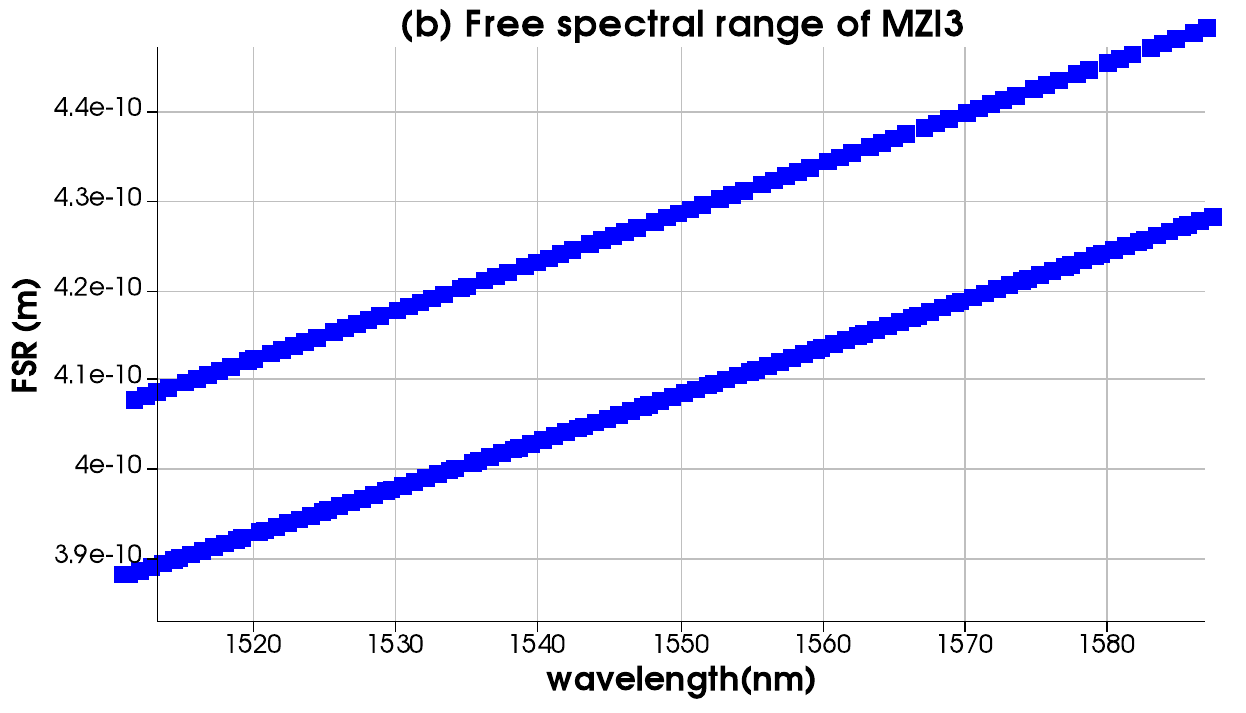}
   (c)\includegraphics[width=0.4\linewidth]{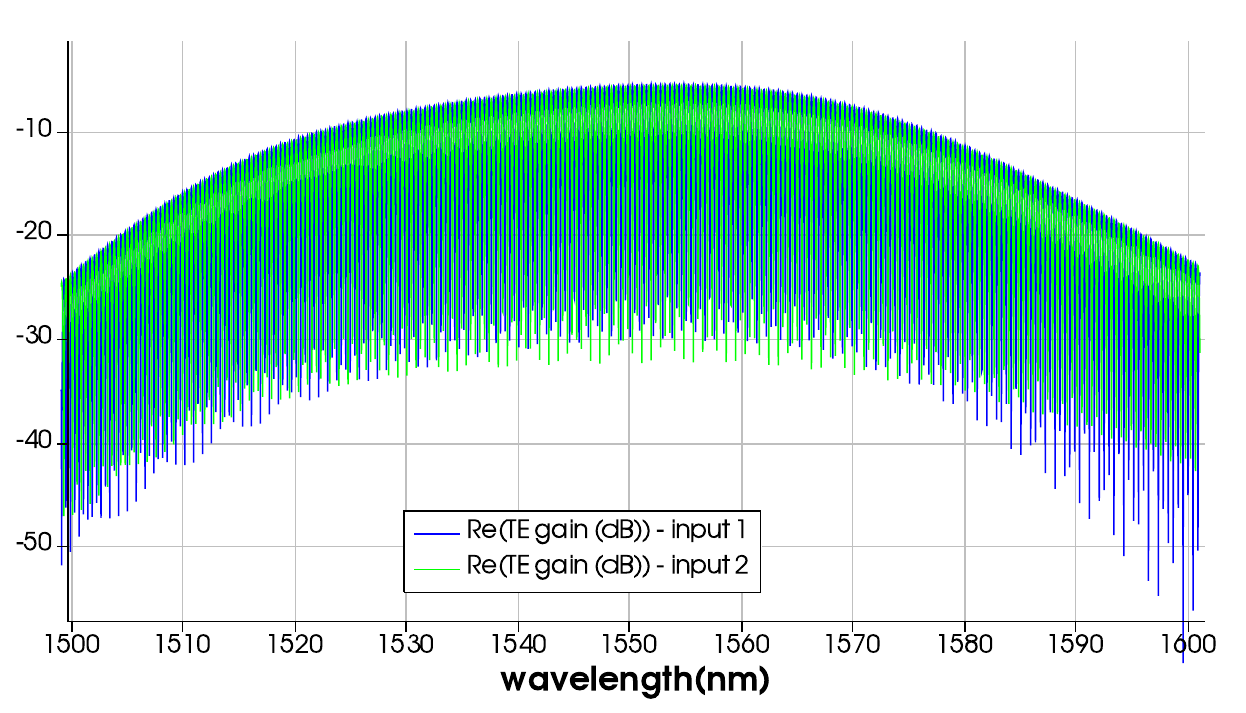}
    (d)\includegraphics[width=0.4\linewidth]{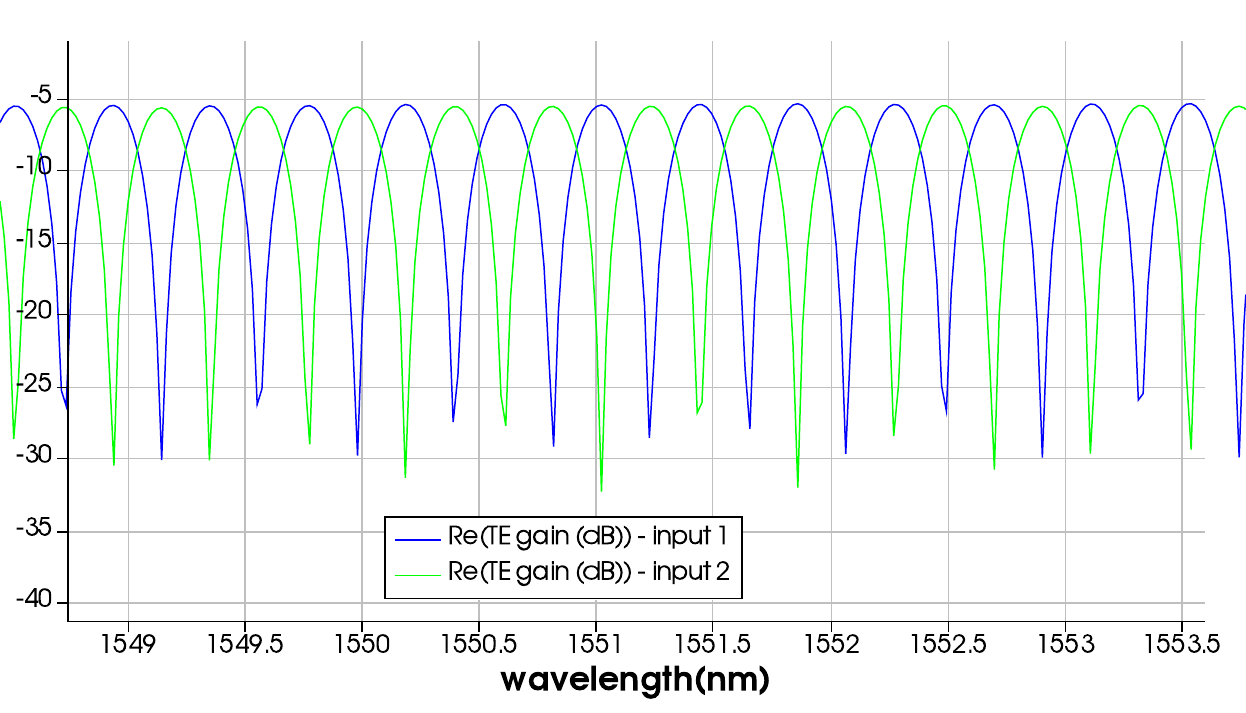}
    
    \caption{The free spectral range (FSR) is less than 1 nm, ranging from 0.41 to 0.45 nm in simulation. The custom MZI (MZI3) transmission spectrum as a function of wavelength. The response is broadband, with narrow peaks and valleys at the sub-nanometer range. However, this type of performance may not necessarily occur in experiment due to manufacturing variability.  }
    \label{fig:7}
\end{figure}

FSR is decreased by extension of path length $L_2$. When $L_2$ increases, the FSR decreases. Although each FSR is less than 50 \si{nm}, the longest difference in waveguide lengths (1373.880 \si{\micro m}) exhibits the narrowest FSR. At the end of the wave-guides, a transverse-electric (TE) broadband splitter combines then separates the signals for the standard configuration and spectrum analysis. Fig. \ref{fig:7} illustrates the transmission spectrum and FSR simulated in ANSYS Optics INTERCONNECT, respectively. Simulated spectra for each of the MZIs is plotted with respect to measured transmission spectra for each MZI.

\section{Results and Discussion}
MZI spectra are measured using an optical spectrum analyzer (OSA), and compared with simulated MZI spectra.  MZIs exhibit differences between measured and simulated spectra by a uniform difference in wavelength, associated with bend losses and effective index. Measured MZI spectra are first compared with simulated MZI spectra from ANSYS Lumerical INTERCONNECT (Supplementary). Least-squares regression fitting is used for each MZI spectrum, since FSR varies with wavelength.

\subsection{Narrow-spectrum Mach-Zehnder Interferometer}

\begin{figure*}
    \centering
    \includegraphics[width=0.32\linewidth]{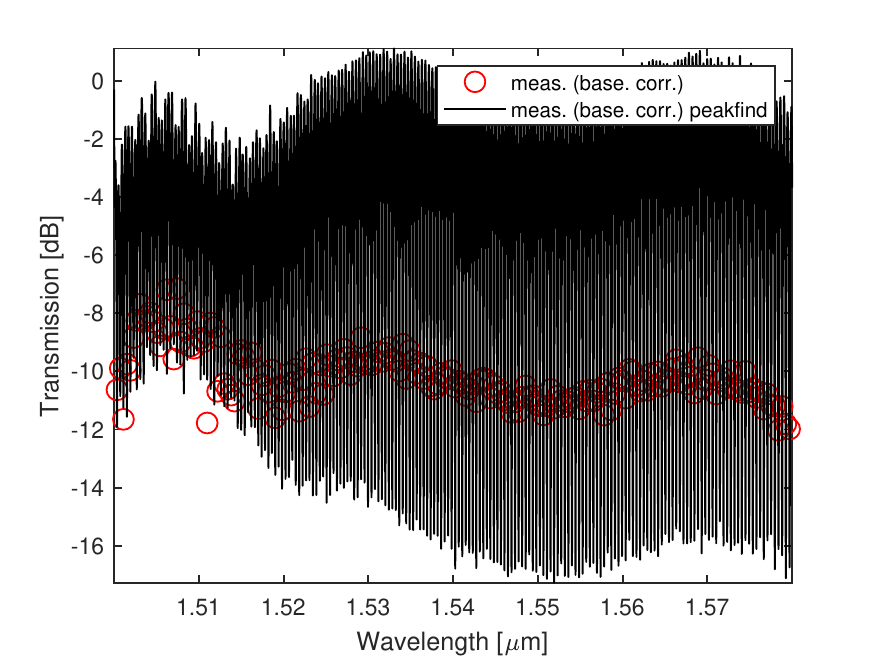}
    \includegraphics[width=0.32\linewidth]{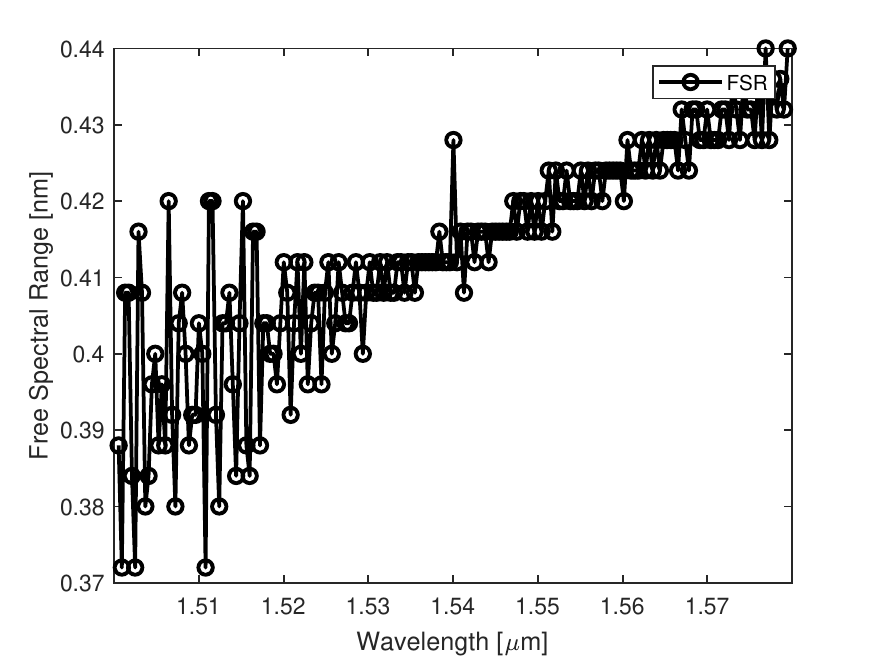} 
     \includegraphics[width=0.32\linewidth]{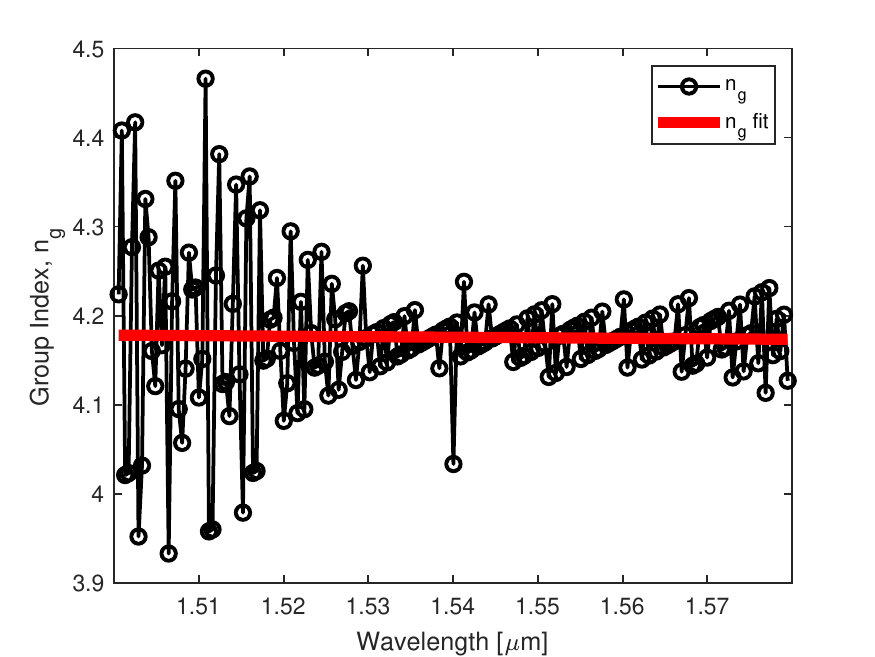} \\
    \includegraphics[width=0.32\linewidth]{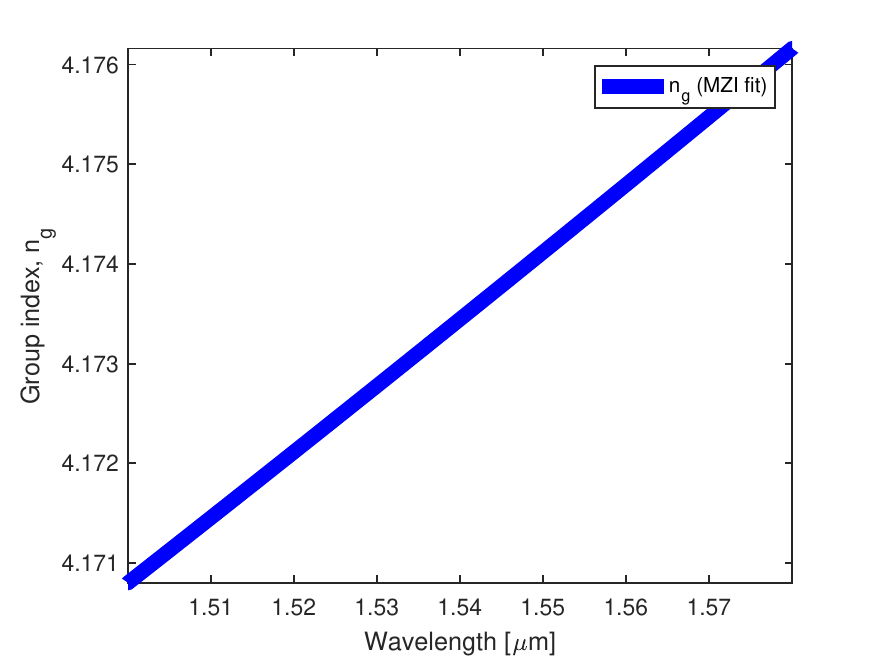}
    \includegraphics[width=0.32\linewidth]{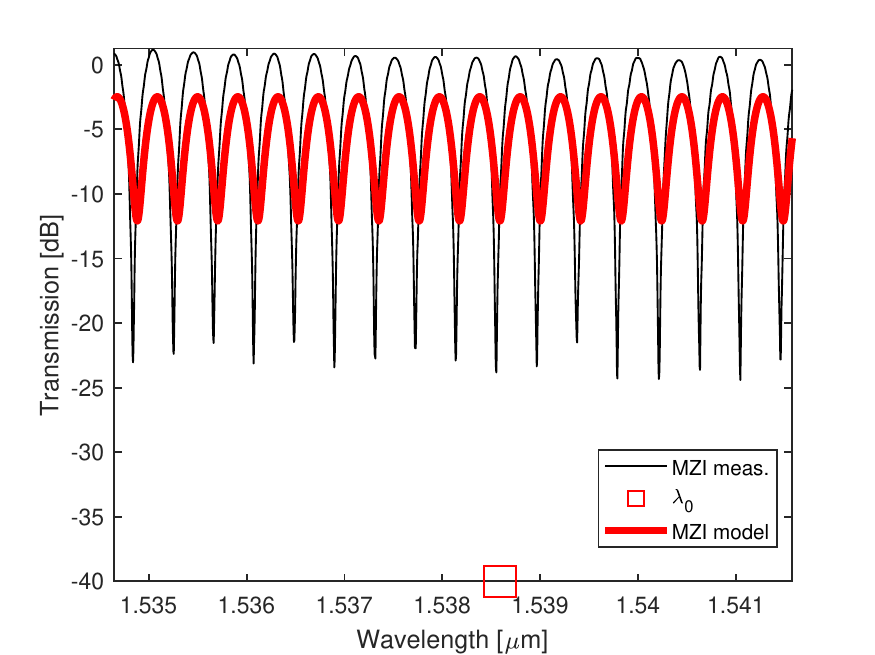}
    \includegraphics[width=0.32\linewidth]{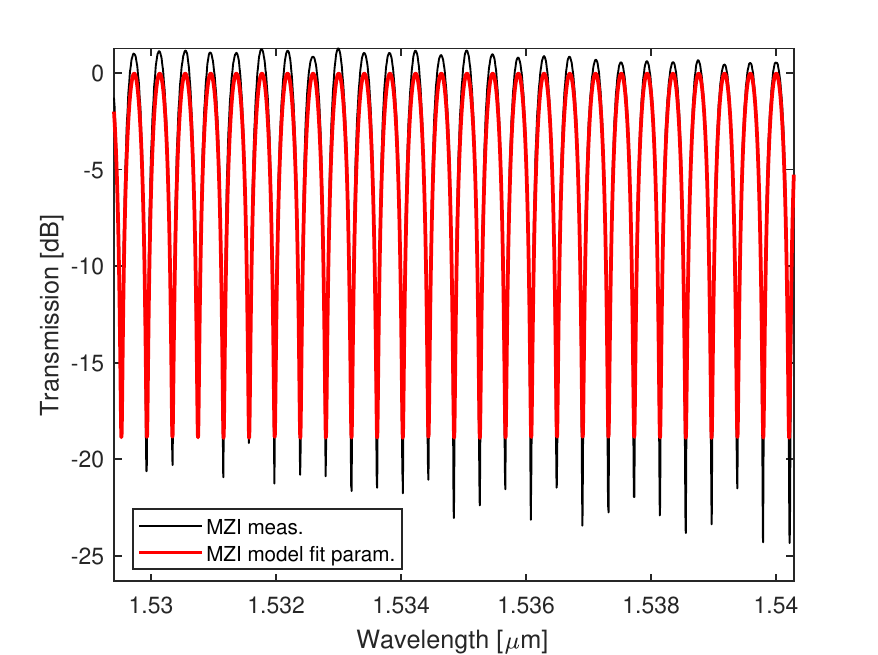} 
    \caption{The fitting process and corresponding model of the narrow FSR MZI TE mode transmission according to Lukas Christowki's method using the \texttt{findpeaks()} and \texttt{polyval()} functions in MATLAB. Group refractive index is recovered within reasonable error tolerance from the default initial condition, such that $n_g=4.1734\pm0.00260$. Similar fits were performed for the remaining MZIs, which recovered $n_g$ values listed in Table 3.}
    \label{fig:10}
\end{figure*}

Fig. \ref{fig:10} illustrates the fitting of the narrow FSR transmission spectrum. Additional losses from dispersion occur outside the wavelength range of $\lambda= 1550\pm5$ \si{nm}.  Good estimates of the baseline are made using a polynomial fit to the local FSR, to recover the group refractive index variation as shown in Fig. \ref{fig:10}. Secondary set of MZIs is considered which have greater FSR, bandwidth, and a similar de-embedding structure to measure process variation.
\begin{figure*}
    \centering
    \includegraphics[width=\linewidth]{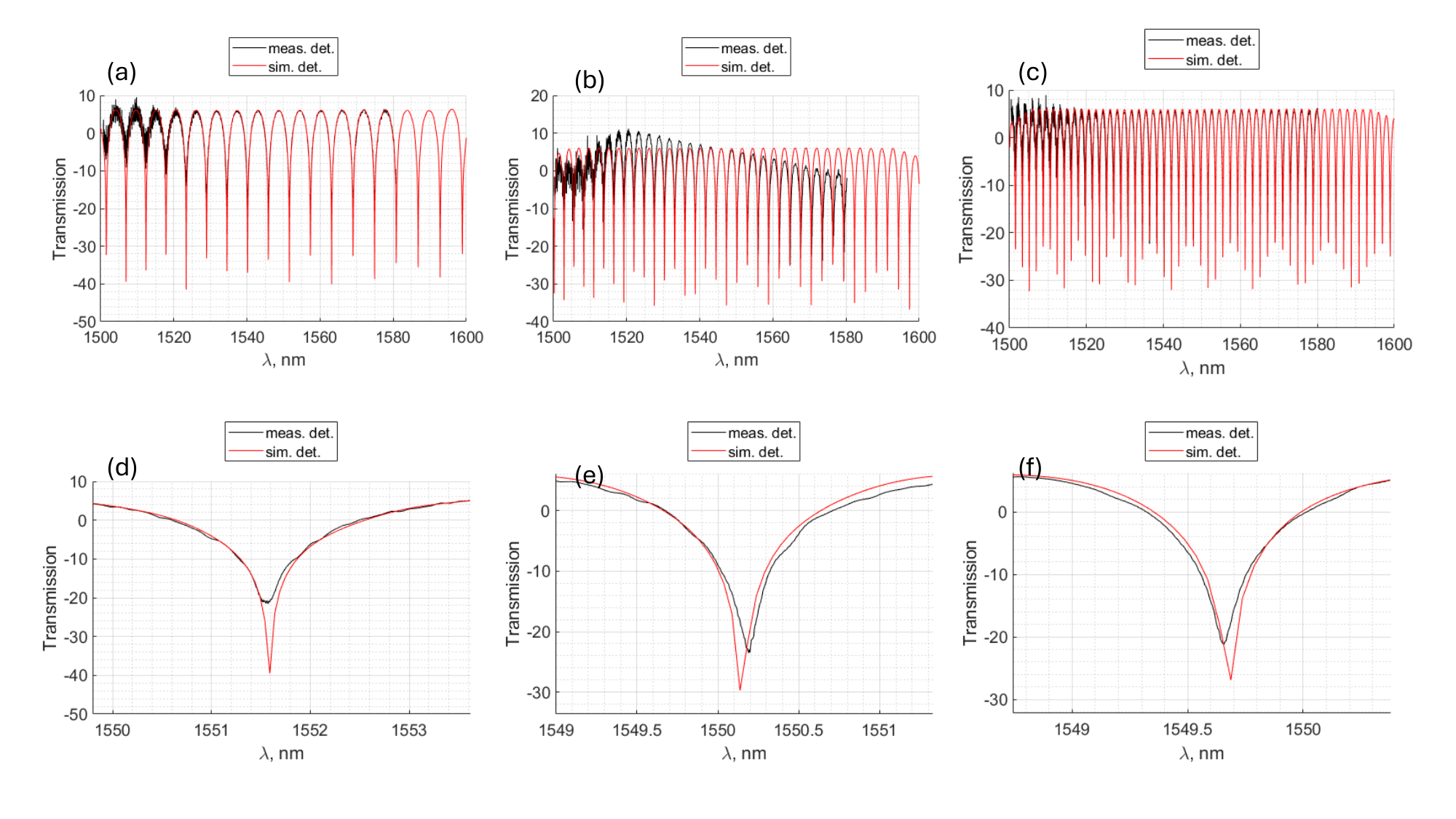}
    \caption{(a) MZI1\_2, (b) MZI2\_2, and (c) MZI3\_2 spectra shifted by wavelength differences $\Delta \lambda=0.71$\si{nm}, $\Delta \lambda=0.23$\si{nm}, and $\Delta \lambda=0.39$\si{nm}, respectively. Central peak alignment of the spectra for (d) MZI1\_2, (e) MZI2\_2, and (f) MZI3\_2. Given the apparent discrepancy from the previous figure, the measurement error is not independent of fitting for process variation error. Therefore, the potential process variation of these MZIs is unknown.}
    \label{fig:13}
\end{figure*}

\begin{table}
    \centering\scriptsize
    \caption{Performance of the MZIs characterized}
    \begin{tabularx}{\textwidth}{|X|c|c|c|c|c|c|c|c|c|}
    \hline
         Component & $\Delta L$ \si{\micro m} & $\mathrm{FSR}$, \si{nm} & $\lambda_0$, \si{nm} & $D$, \si{ps/nm/km} & $n_g$ & $\mathrm{R}^2$ & $n_1$ & $n_2$ & $n_3$ \\
         \hline
         NMZI (MZI3\_1) &1373.880&0.41325$\pm$0.01418&1538.6&223.3075&4.1734$\pm$0.00260&0.91591&2.39935&-1.15457&0.01891\\
         MZI1\_2&100.000&5.6529$\pm$0.16468&1533.8&252.6655&4.1830$\pm$0.00252&0.90531&2.38506&-1.17031&-0.14519\\
         MZI2\_2&200.000&2.8264$\pm$0.10826&1535.7&691.3808&4.1759$\pm$0.00728&0.87603&2.39961&-1.16740&0.18163\\
         MZI3\_2&300.000&1.8896$\pm$0.06076&1537.9&256.5658&4.1777$\pm$0.00284&0.91270&2.39655&-1.16014&0.01613 \\
         MZI4\_1&150.000&3.786$\pm$0.1276&1535.3&115.4766&4.1768$\pm$0.00125&0.90752&2.38998&-1.16278&-0.09261 \\
         MZI5\_1&250.000&2.265$\pm$0.07512&1536.0&403.2915&4.1772$\pm$0.00420&0.91199&2.39921&-1.16043&0.04330 \\
         MZI6\_1&350.000&1.6209$\pm$0.05768&1538.2&203.7255&4.1771$\pm$0.00234&0.92538&2.39735&-1.15847&0.00905 \\
         \hline
    \end{tabularx}
    \label{tab:3}
\end{table}

\subsection{Variable path-length difference of broad spectrum Mach-Zehnder interferometer}
In this design, the same procedures were used as for the narrow spectrum MZI, except that the path length is varied and the FSR increased for a consistent measurement. Spectral fitting is performed using a wavelength shift and the MATLAB \texttt{findpeaks()} function \cite{ref2}. Following a wavelength shift, each spectrum shares excellent agreement with ANSYS Optics INTERCONNECT simulated spectra within 0.1 \si{nm} (Fig. \ref{fig:13}). Therefore, the difference from the measurement is attributed to effective index, whereas the group index is similar between the designs with minor variation.

Path length differences for each of these MZI are smaller than in the narrow spectrum MZI, and less prone to process variation. MZI1\_2, MZI2\_2 and MZI3\_2 exhibit the greatest wavelength shift, which is uncorrelated to their path lengths. Wavelength shifts in MZI4\_1, MZI5\_1 and MZI6\_1 is negligible, suggesting variation in effective index was a minor effect for these waveguides which share an excellent agreement with the simulated spectra (Fig. \ref{fig:13}). In this chip, a de-embedding structure was also measured for a waveguide connected directly to fiber grating couplers, y-branches and terminator branches (YWGD structure). After performing baseline subtraction of the YWGD spectrum from each MZI transmission spectrum, variation in transmission from the MZI structures also emphasize variation in insertion loss. However, it is questionable whether process variation occurs between the de-embedding structure and the MZI, such that a polynomial fit is more adequate for analysis and fitting of the spectrum.

\subsection{Fitted group refractive index and dispersion}
\begin{figure}[H]
    \centering
\includegraphics[width=0.32\linewidth]{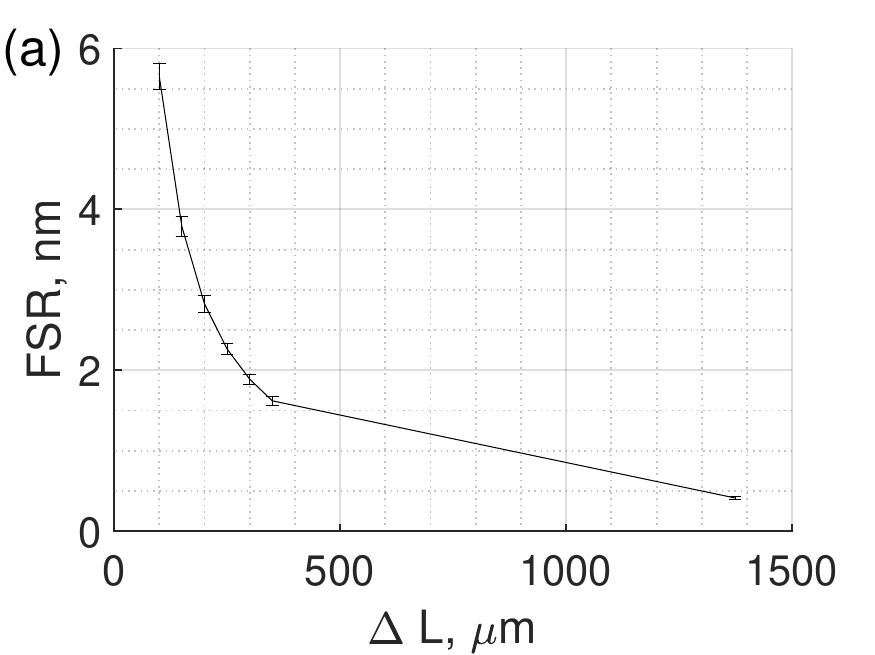}
\includegraphics[width=0.32\linewidth]{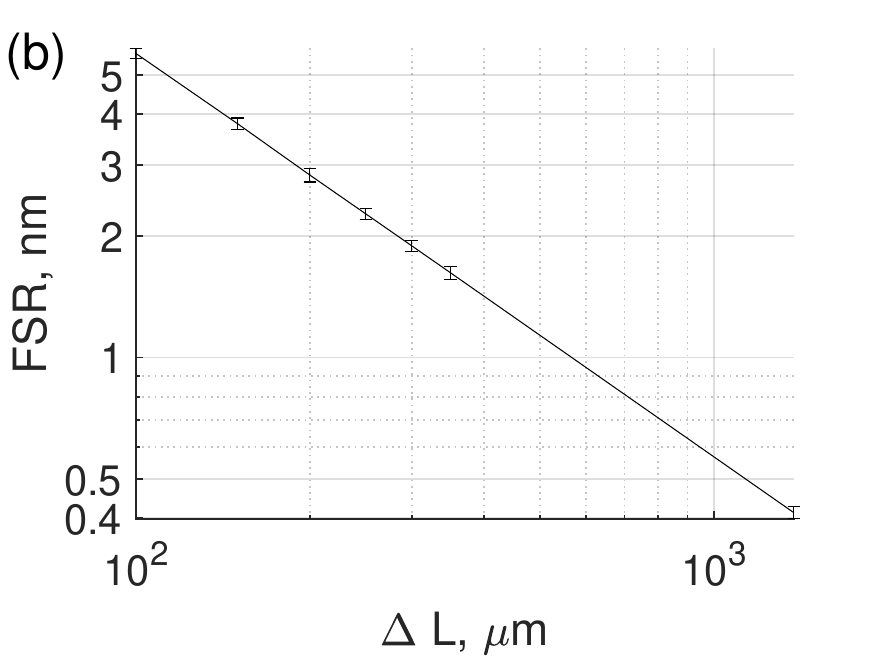}
(c)\includegraphics[width=0.32\linewidth]{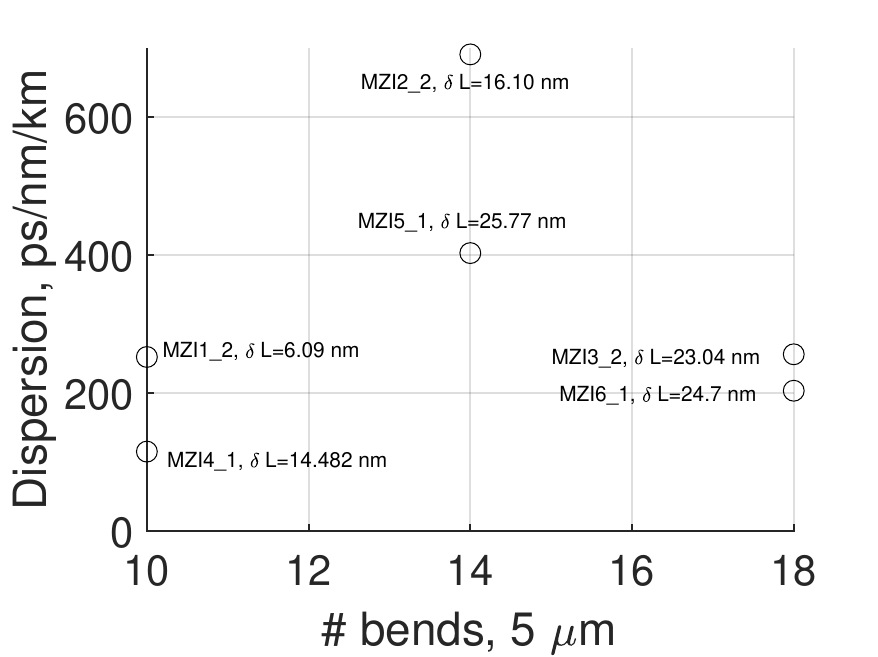}
    \caption{Measured free-spectral range (FSR) and their deviations for each of Mach-Zehnder interferometer designs, as a function of their path length difference, $\Delta L$, on a (a) linear scale and a (b) log-log scale. The greater the path length difference, the smaller the FSR. Deviation in FSR is uniform with respect to the FSR. (c) Dispersion, $D$, decreases as the linear interbend path length, $\delta L$, increases for the same number of bends with a similar geometric configuration. The cause for this is attributed to linear waveguide enhanced single mode coupling. }
    \label{fig:14}
\end{figure}

The fitted group refractive index and dispersion for each of the MZI components is listed in Table \ref{tab:3}. Note that the narrow FSR MZI (NMZI or MZI3\_1) has similar dispersion as MZI1\_2. Therefore, all  measured MZIs perform nominally, albeit with varying group refractive index and dispersion. Relative differences in dispersion with interbend path length, $\delta L$, for the same number of bends is illustrated in Fig. \ref{fig:14}c. The lowest dispersion is obtained from MZI4\_1 ($D=115.4766$ \si{ps/nm/km}), which has a greater path length than MZI1\_2 ($D=252.6655$ \si{ps/nm/km}), but with the same number of bends. THerefore, a shorter linear path length between consecutive bends increases dispersion more than a longer linear path length between consecutive bends. Similarly, although MZI5\_1 ($D=403.2915$ \si{ps/nm/km}) and MZI6\_1 ($D=203.7255$ \si{ps/nm/km}) have similar number of bends as MZI2\_2 ($D=691.3808$ \si{ps/nm/km}) and MZI3\_2 ($D=256.5658$ \si{ps/nm/km}), respectively, their increased path lengths between consecutive bends result in less dispersion. These observations explain why NMZI (MZI3\_1) achieves lower dispersion than MZI1\_2, despite a lower FSR, greater path length, and greater number of bends.   

\section{Conclusion}
Mach-Zehnder interferometers can achieve a low free spectral range to sub-nanometer resolution using a longer path difference between each branch (EBeam\_cael3.gds), with longer linear misaligned wave-guide sections between consecutive bends with optimally short bend radius. These strategies may help minimize orbital drift of the electron beam during the fabrication process, and avoid modal dispersion from stepped electron beam machine tolerance at wider bends. Dispersion is reduced in longer sections between consecutive bends due to enhanced single mode coupling, which occurs naturally in long optical fibers and waveguides. Although losses are increased due to the path length, the ANSYS Optics INTERCONNECT simulated transmission spectrum and the measured transmission spectrum are perfectly aligned near the center wavelength by adding a negative constant, suggesting the cause as insertion loss. Long linear waveguides also decrease fabrication cost, by decreasing tool time with faster operation of the electron beam.

In the case of MZIs with varying path length difference, the dispersion manifests as a wavelength shift in each spectrum which is uncorrelated to the path length difference. Accounting for dispersion, the measured transmission spectra are similar to the simulated spectra in ANSYS Optics INTERCONNECT independent of wavelength (within $\mathrm{err}(\lambda)<0.1$ \si{nm} for each MZI). Fitting spectra with least squares regression of a Taylor series approximation for $n_{eff}$ and wavelength dependent FSR reveals increasing dispersion associated with decreasing path length between consecutive bends. Excellent agreement between the measured and simulated spectra is obtained using minimum radius bends with increased linear separation distance.

\section{Acknowledgments}
    I acknowledge the edX UBCx Phot1x Silicon Photonics Design, Fabrication and Data Analysis course, which is supported by the Natural Sciences and Engineering Research Council of Canada (NSERC) Silicon Electronic-Photonic Integrated Circuits (SiEPIC) Program. The devices were fabricated by Cameron Horvath at Applied Nanotools, Inc. Omid Esmaeeli performed the measurements at The University of British Columbia. We acknowledge Lumerical Solutions, Inc., Mathworks, Python, and KLayout for the design software.  This work was also supported in part by NSERC CGSD and Alberta Innovates grants.


\begin{thebibliography}{1}
\bibitem{ref1}
R. J. Bojko, J. Li, L. He, T. Baehr-Jones, M. Hochberg, and Y. Aida, ``Electron beam lithography writing strategies for low loss, high confinement silicon optical waveguides," J. Vacuum Sci. Technol. B 29, 06F309 (2011)
\bibitem{ref2}
L. Chrostowski and M. Hochberg. {\it{Silicon photonics design: from devices to systems.}} Cambridge University Press, 2015.
\bibitem{ref3}
\href{http://siepic.ubc.ca/probestation}{http://siepic.ubc.ca/probestation}, using Python code developed by Michael Caverley.
\bibitem{ref4}
Yun Wang, Xu Wang, Jonas Flueckiger, Han Yun, Wei Shi, Richard Bojko, Nicolas A. F. Jaeger, Lukas Chrostowski, "Focusing sub-wavelength grating couplers with low back reflections for rapid prototyping of silicon photonic circuits", Optics Express Vol. 22, Issue 17, pp. 20652-20662 (2014) doi: 10.1364/OE.22.020652
\bibitem{ref5}
\href{www.plcconnections.com}{www.plcconnections.com}, PLC Connections, Columbus OH, USA.
\bibitem{ref6}
\href{http://mapleleafphotonics.com}{http://mapleleafphotonics.com}, Maple Leaf Photonics, Seattle WA, USA.
\end{thebibliography}
\end{document}


\title{Supplementary document for ``Low-dispersion low free-spectral-range Mach-Zehnder interferometer with long straight path lengths on silicon"}
\author{\href{https://orcid.org/0000-0003-3493-527X}{\includegraphics[scale=0.06]{orcid.pdf}\hspace{1mm}Cael~Warner}\thanks{Electrical and Computer Engineering\\
University of Alberta\\
116 St \& 85 Ave, Edmonton T6G 2R3 \\
\texttt{spencerw@ualberta.ca} 
}}

\renewcommand{\shorttitle}{Low FSR MZI on Silicon, Supplementary}

\hypersetup{
pdftitle={Low dispersion low FSR MZI long path lengths Si supplementary},
pdfsubject={},
pdfauthor={Cael~Warner},
pdfkeywords={Mach-Zehnder, interferometer, silicon},
}

\maketitle

\begin{abstract}
This supplementary material describes the detailed device fabrication. Additionally, it illustrates the simulated modes at the Y-branch, radius bends, and the fitting process for the spectra to determine the free spectral range, group index, effective index, and dispersion.
\end{abstract}

\keywords{Mach-Zehnder \and interferometer \and silicon}

\section{SUMMARY OF THE FABRICATION DESCRIPTION}
Fabrication is performed at one or more of these: Applied Nanotools and Washington Nanofabrication Facility. The following are the process descriptions.

\subsection{Applied Nanotools, Inc. NanoSOI process}
The MZI structures are part of a NanoSOI multi-process wafer fabrication by Applied Nanotools Inc. (\href{http://www.appliednt.com/nanosoi}{http://www.appliednt.com/nanosoi}; Edmonton, Canada) based on direct-write 100 \si{keV} electron beam lithography technology. The silicon-on-insulator wafer has 220 \si{nm} thickness and a 2 \si{\micro m} buffer oxide thickness as the base material. The wafer was initially cleaned using piranha solution (3:1 H\textsubscript{2}SO\textsubscript{4}:H\textsubscript{2}O\textsubscript{2}) for 15 minutes and water/IPA rinse, then hydrogen silsesquioxane (HSQ) resist was spin-coated onto the substrate which is heated to vaporize the solvent. Each MZI is patterned using a JEOL JBX-8100FS electron beam instrument at The University of British Columbia (UBC). The exposure dosage was corrected to account for proximity backscatter of electrons from nearby features. Shape writing order was optimized for minimal beam drift. After the e-beam exposure and development with tetramethylammonium sulfate (TMAH) solution, the devices are inspected for residues and/or defects. The chips were mounted on a 4” handle wafer for anisotropic ICP-RIE etch using chlorine with a qualified etch rate. Photoresist is removed from the surface of each MZI using a 10:1 buffer oxide wet etch, and the devices inspected using a scanning electron microscope (SEM) to verify patterning and etch quality. A 2.2 \si{\micro m} oxide cladding is deposited using plasma-enhanced chemical vapour deposition (PECVD) based on tetraethyl orthosilicate (TEOS) at 300$^\circ$C. Reflectrometry measurements performed throughout the process verified the device layer, buffer oxide and cladding thickness.

\section{Waveguide modes}
Mode confinement for the standard strip waveguide geometry ($w=500$ \si{nm}, $h=200$ \si{nm}), is simulated in ANSYS Optics MODE Eigenmode solver for the transverse electric (TE) field configuration. 
\subsection{Mode confinement in linear waveguide}
Modes are solved for near the refractive index for silicon. The TE mode is confined to the center of the waveguide, whereas the TM mode is confined to the periphery of the waveguide. Changing the width and height of the standard waveguide structure in the Eigenmode solver realizes different eigenfrequencies, mode confinement, group index, and effective index, which can be used in a corner analysis to determine their effect on the transmitted spectra. 

\begin{figure}[H]
    \centering
    \includegraphics[width=\linewidth]{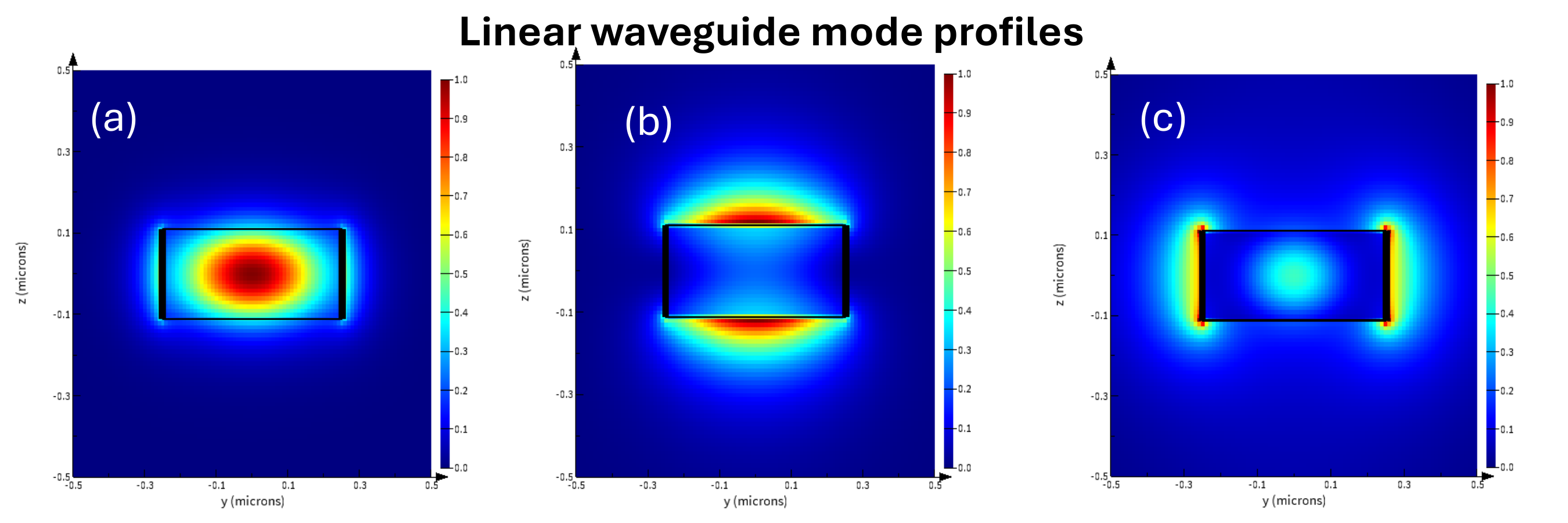}
    \caption{(a) TE Mode; (b) TM Mode; and (c) second order TM mode. (c) Will coupled to a longer waveguide, whereas only (a) and (b) are the desired modes.}
    \label{fig:4}
\end{figure}

\subsection{Mode confinement in bend and curvilinear waveguide}
Modes are solved for near the refractive index for silicon with a standard bend radius of 5 \si{\micro m}. In this case, there is only one TE and TM mode recovered by the solver, despite the strong displacement of the mode from the center of the waveguide. It requires an extended path length for the modal dispersion to evolve. In the case of a longer waveguide bend radius, such as 20 \si{\micro m}, an additional mode appears with significant losses into the silicon dioxide layer. This illustrates the effect of increased path length bends on the dispersion of modes confined modes in the strip waveguide. Although a longer bend radius may reduce transmission losses for the principal mode overall, it has the undesired modal dispersion. Therefore, the final design uses minimum bend radii. The limitation of the minimum bend radius is variation in effective refractive index from bend losses.

\begin{figure}[H]
    \centering
    \includegraphics[width=\linewidth]{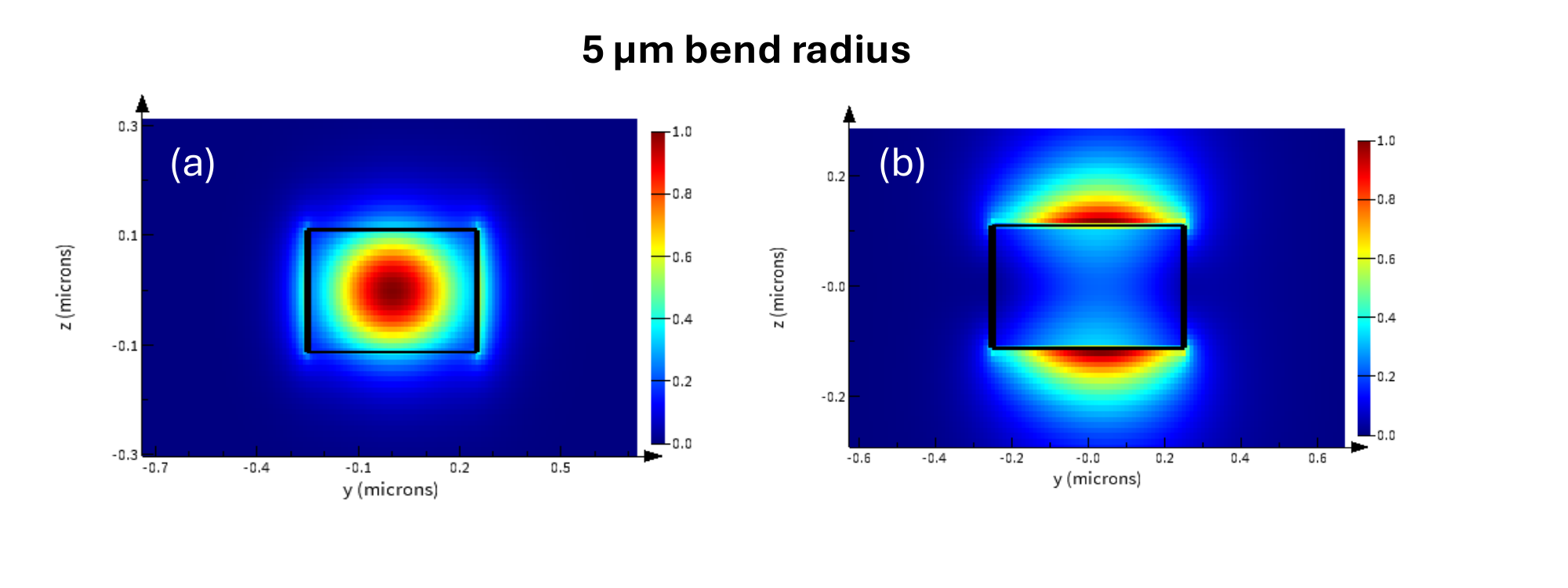} \\
    \includegraphics[width=\linewidth]{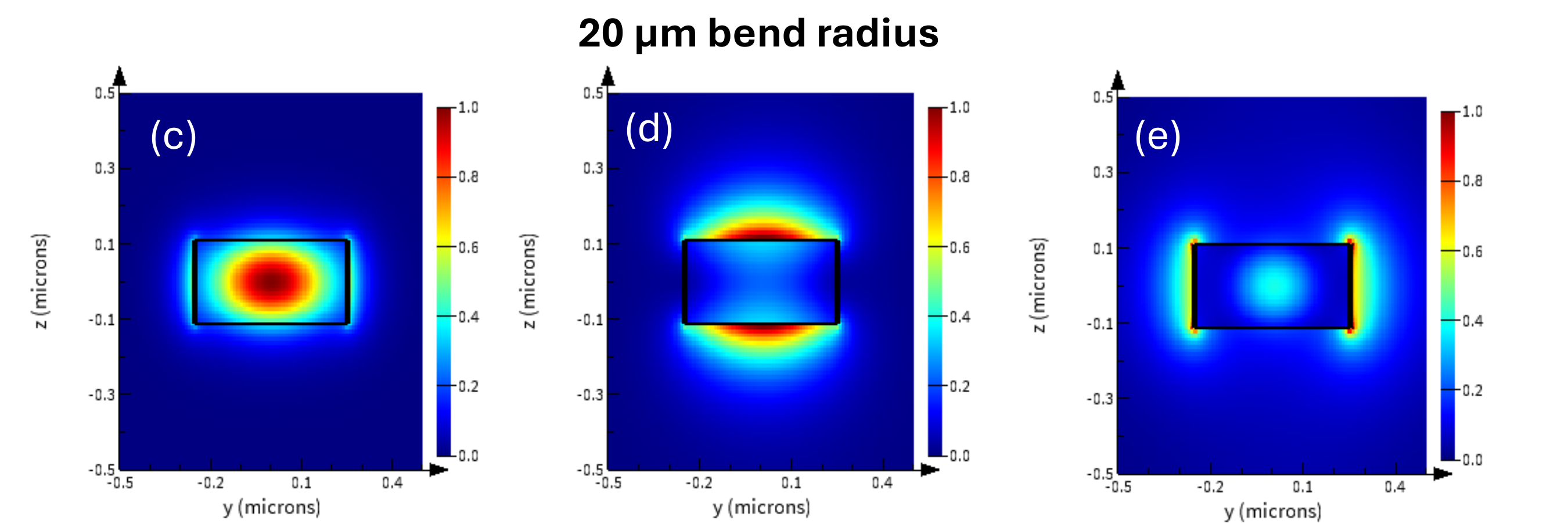}
    \caption{(a) TE mode and (b) TM mode for the 5 \si{\micro m} bend radius. (c) TE mode; (b) TM mode; and (c) second order dispersive TM mode for the 20 \si{\micro m} bend radius. Increasing the bend radius increased the dispersion of the second-order dispersive TM mode, which is associated with greater loss.}
    \label{fig:5}
\end{figure}

\section{Y-branch losses}
\begin{figure}[H]
    \centering
    \includegraphics[width=\linewidth]{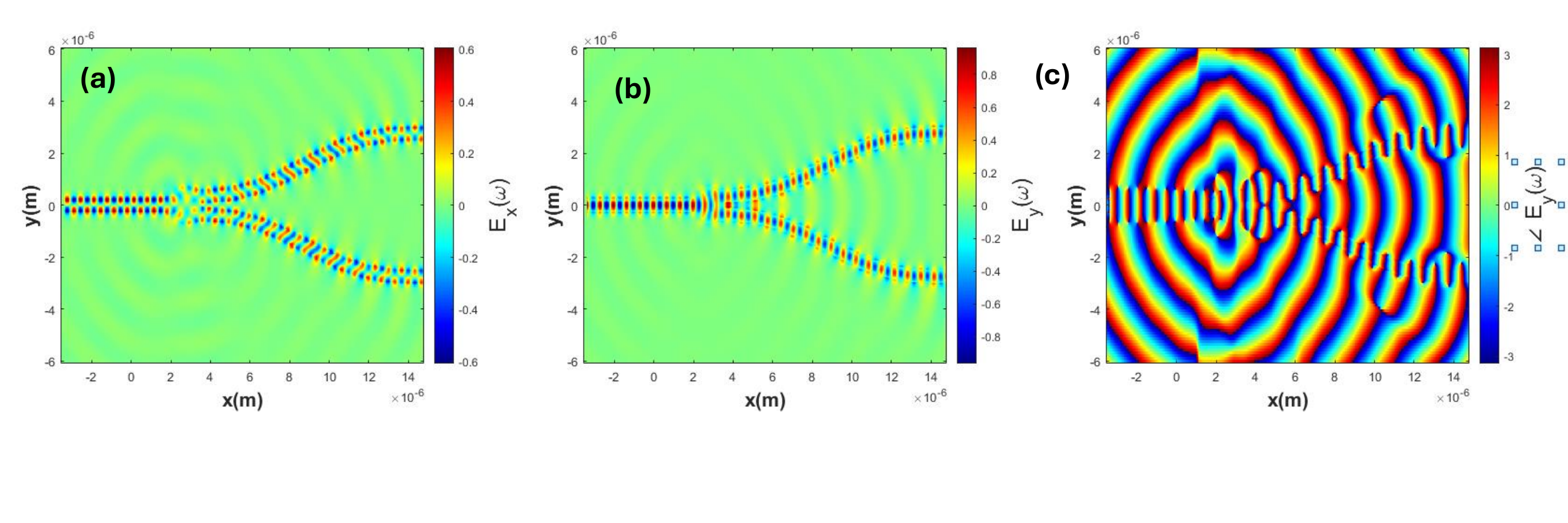}
    \caption{The Y-branch confined frequency-domain electric field components $E_x(\omega)$ in (a) and $E_y(\omega)$ in (b). The phase of $E_y(\omega)$ in (c). Note that as the wave passes the Y-branch node, modal dispersion results. The phase that originally enters the Y-branch is not recovered, as illustrated in (c). Radiating modes enter the silicon dioxide. These radiating modes easily couple with adjacent elements if they are too near.}
    \label{fig:6}
\end{figure}

The Y-branch losses are simulated in ANSYS Optics varFDTD as a 2.5 FDTD solver. Since individual Y-branches are not measured in the experiment, it is not possible to use simulated loss from the Y-branch directly. However, the S-parameters associated with the guided modes can be used between consecutive custom elements in order to predict the performance of connected devices in ANSYS Optics INTERCONNECT. Despite performing these simulations in the course, they were not necessary given the existing components in the UBC EBeam PDK. Figure \ref{fig:6} illustrates simulated frequency domain, $E_x(\omega)$ and $E_y(\omega)$, field components as they occurred in the waveguide. The corresponding phase for $E_y(\omega)$ should have repeated wavefronts, but after passing through the waveguide, the wavefronts are distorted due to modal dispersion. Numerical dispersion may also occur, so a fine uniform mesh was used to mitigate the artificial modes. Notice that after electromagnetic waves pass through each branch, the transverse electric field has a difference in its phase. Differences in phase here also occur in the measured transmission spectra, and are associated with bend losses but described using an effective refractive index assuming a similar structure. After the Y-branch, the phase in each waveguide approaches the phase of the fundamental mode. This phenomenon may be referred to as single mode coupling, and occurs in optical fibers which have low dispersion losses over long distances. Therefore, it is advantageous to consider increased linear path length between consecutive bends to minimize dispersion losses.

\section{Measurement data analysis}

Curve fitting for baseline correction requires accurate initial conditions to fit with both local and global minima and maxima. In order to inform the curve fitting method, it is important to have an analytical or numerical solution in advance. Waveguide parameters can be extracted from the following curvefit formulae for group index:
\begin{equation}
    n_g=n_{
    \mathrm{eff}}-\lambda\frac{dn_{\mathrm{eff}}}{d\lambda}=n_1-n_2\lambda_0,
\end{equation}
and group velocity dispersion:
\begin{equation}
    D=\frac{\lambda}{c}\frac{d^2}{d\lambda^2} n = -2\frac{\lambda}{c}n_3,
\end{equation}
where $D$ is in units of \si{ps/nm/km}. Given this information, the curve fitting target function for the MZI is 
\begin{equation}
    F=10\log_{10}\left(\frac{1}{4}\left|1+\exp\left[-i\frac{2\pi n_{\mathrm{eff}}}{\lambda}\Delta L - \frac{\alpha\Delta L}{2}\right]\right|\right)+b,
\end{equation}
where $\Delta L$ is the difference in the wave-guide path length between the two branches, $\lambda$ is the wavelength in the wave-guide, $n_\mathrm{eff}$ is the effective refractive index, and $\alpha$ is the linear absorption coefficient, and $b$ is the excess insertion loss. We may solve this provided an expression for effective index from the Taylor series expansion 
\begin{equation}
    n_{\mathrm{eff}}=n_1+n_2(\lambda-\lambda_0)+n_3(\lambda-\lambda_0)^2,
\end{equation}
where $\lambda_0$ is the free space wavelength. Our objective is to find $n_1$, $n_2$, and $n_3$ so that we may recover the effective refractive index for a device that has been tested. In order to accomplish this, least-squares fitting can be used. After fitting, the measured results can be compared with simulated results returned from the eigenmode solver.

MZI fitting of the group index can be solved using the MATLAB \texttt{findpeaks()} function, which can recover the local minima and maxima in the MZI spectrum. The free spectral range
\begin{equation}
    \mathrm{FSR}[m]=\lambda_{n+1}-\lambda_n,
\end{equation}
is determined from these local minima and maxima provided the correct settings in the \texttt{findpeaks()} function. Given the free spectral range, the group refractive index may be recovered provided a known wavelength $\lambda$ and difference in length between the MZI branches, $\Delta L$, as
\begin{equation}
    n_g=\frac{\lambda^2}{\Delta L\mathrm{FSR}[m]}.
\end{equation}
Given $n_g$, we can recover $n_2$. First, however, we must recover $n_1$ by choosing a mid-peak in the spectrum and matching the model at that peak. In this case, mode number $N$ for one of the destructively interfering modes is recovered from:
\begin{equation}
    \cos{\left(\frac{2\pi n_1\Delta L}{\lambda_0}\right)}=-1
\end{equation}
such that
\begin{equation}
    \frac{2n_1\Delta L}{\lambda_0}=2N+1.
\end{equation}
In order to find $N$, we can use an approximate $n_1\approx2.4$, such that
\begin{equation}
    N=\mathrm{Round}\left(\frac{n_1\Delta L}{\lambda_0}-\frac{1}{2}\right).
\end{equation}
Provided the initial $n_\mathrm{eff}$ expression, $n_g$, and $n_1$, the group index can be estimated by substituting the Taylor series expansion of (4) into  (1), such that
\begin{equation}
    n_2=-\frac{n_g-n_1}{\lambda_0}.
\end{equation}
Finally, we can determine a curve fitting between the group index and the wavelength. The slope of the fitted curve is directly proportional to $n_3$, 
\begin{equation}
    n_3=-\frac{1}{2\lambda}\frac{d n_g}{d \lambda}
\end{equation}
which can be used in the initial condition for the curve fitting target function (3).

\begin{figure}[H]
    \centering    \includegraphics[width=0.32\linewidth]{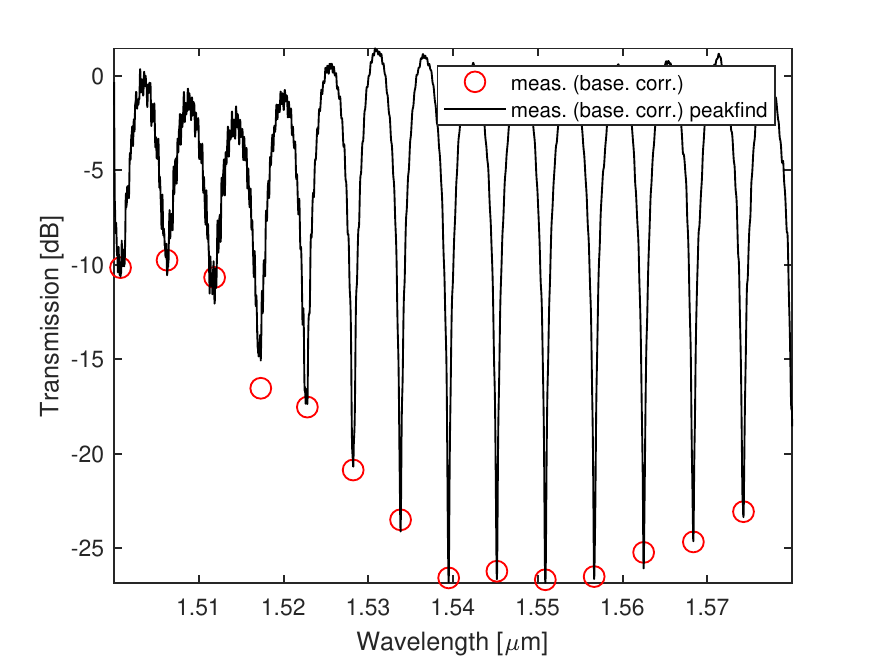}
    \includegraphics[width=0.32\linewidth]{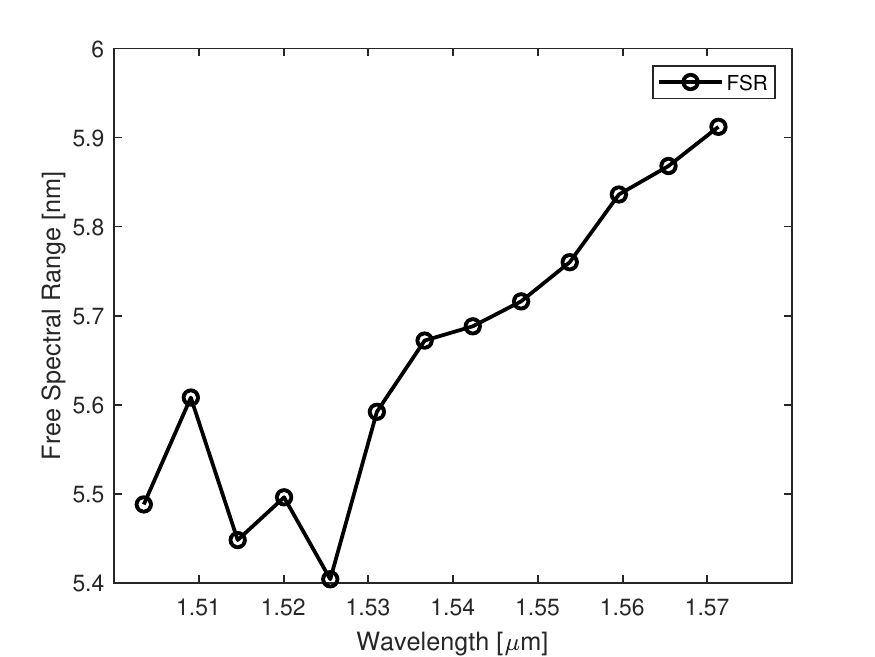}
     \includegraphics[width=0.32\linewidth]{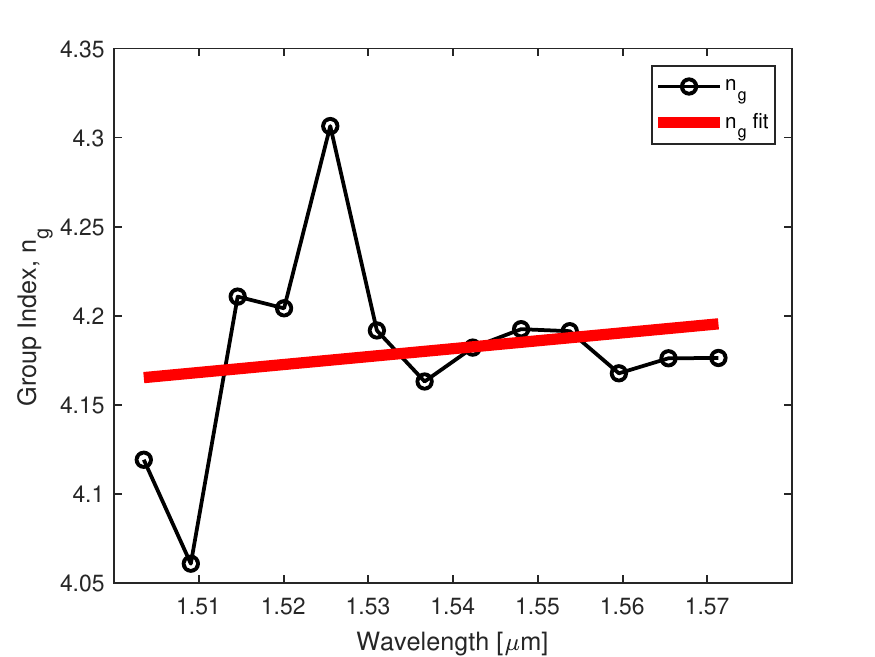}
    \includegraphics[width=0.32\linewidth]{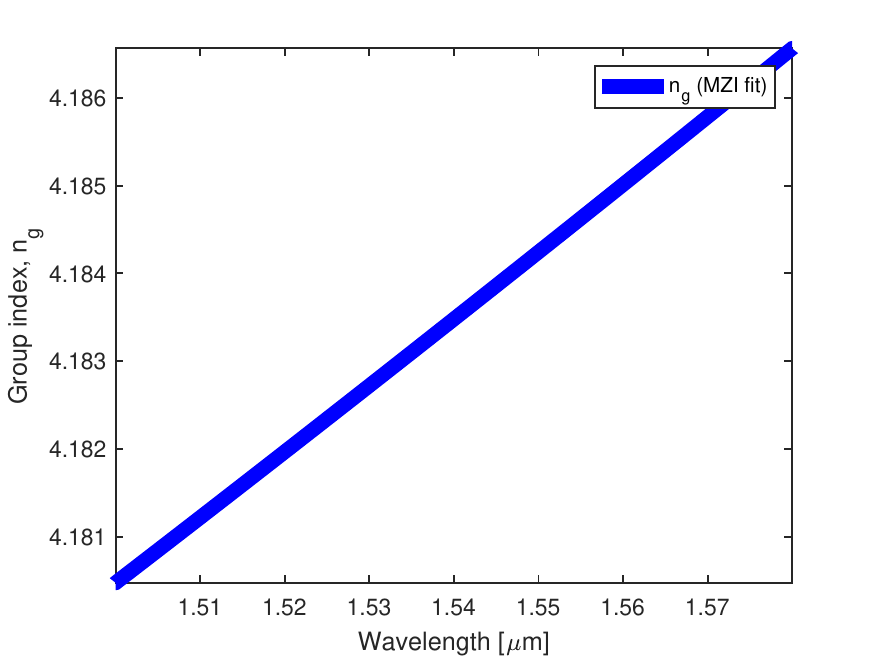}
    \includegraphics[width=0.32\linewidth]{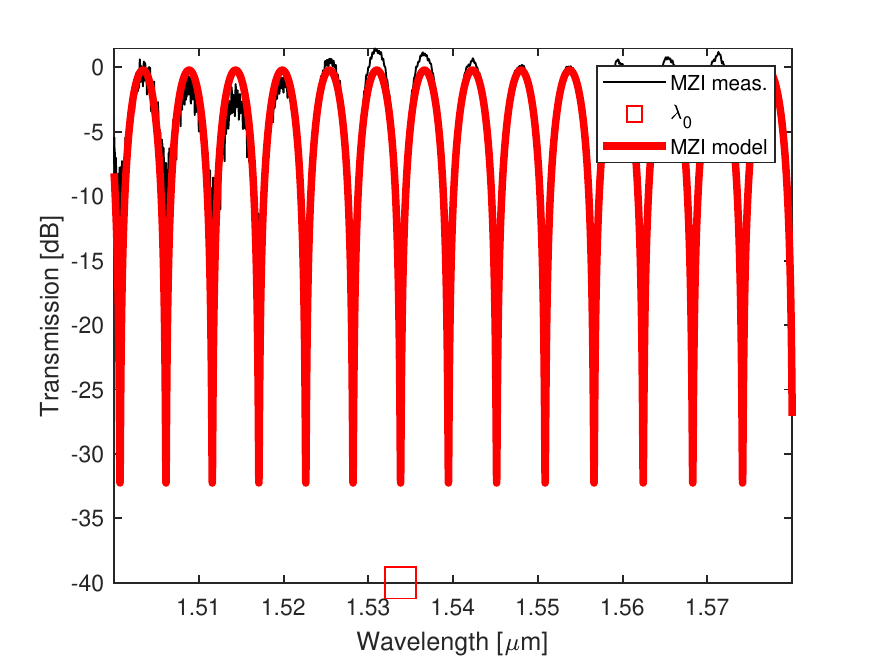}
    \includegraphics[width=0.32\linewidth]{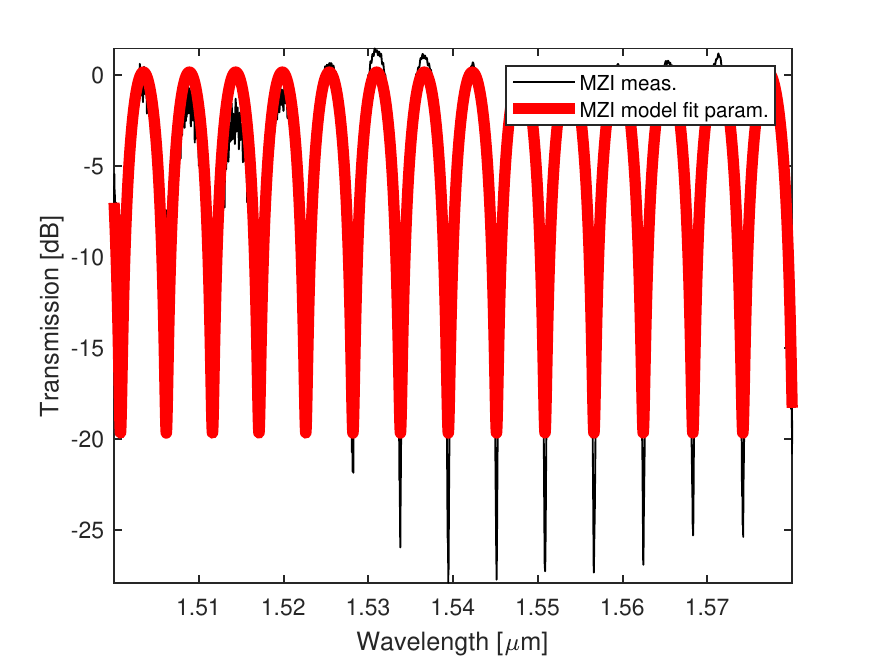} 
    \caption{The fitting process and corresponding model of MZI1\_2 TE mode transmission (EBeam\_cael3\_v2.gds) according to Lukas Christowki's method using the \texttt{findpeaks()} and \texttt{polyval()} functions in MATLAB. Group refractive index is recovered within reasonable error tolerance from the default initial condition, such that $n_g=4.18301\pm0.00250$. Similar fits were performed for the remaining MZIs, which recovered $n_g$ values listed in Table 3.}
    \label{fig:12}
\end{figure}

\begin{figure}[H]
    \centering
    \includegraphics[width=0.49\linewidth]{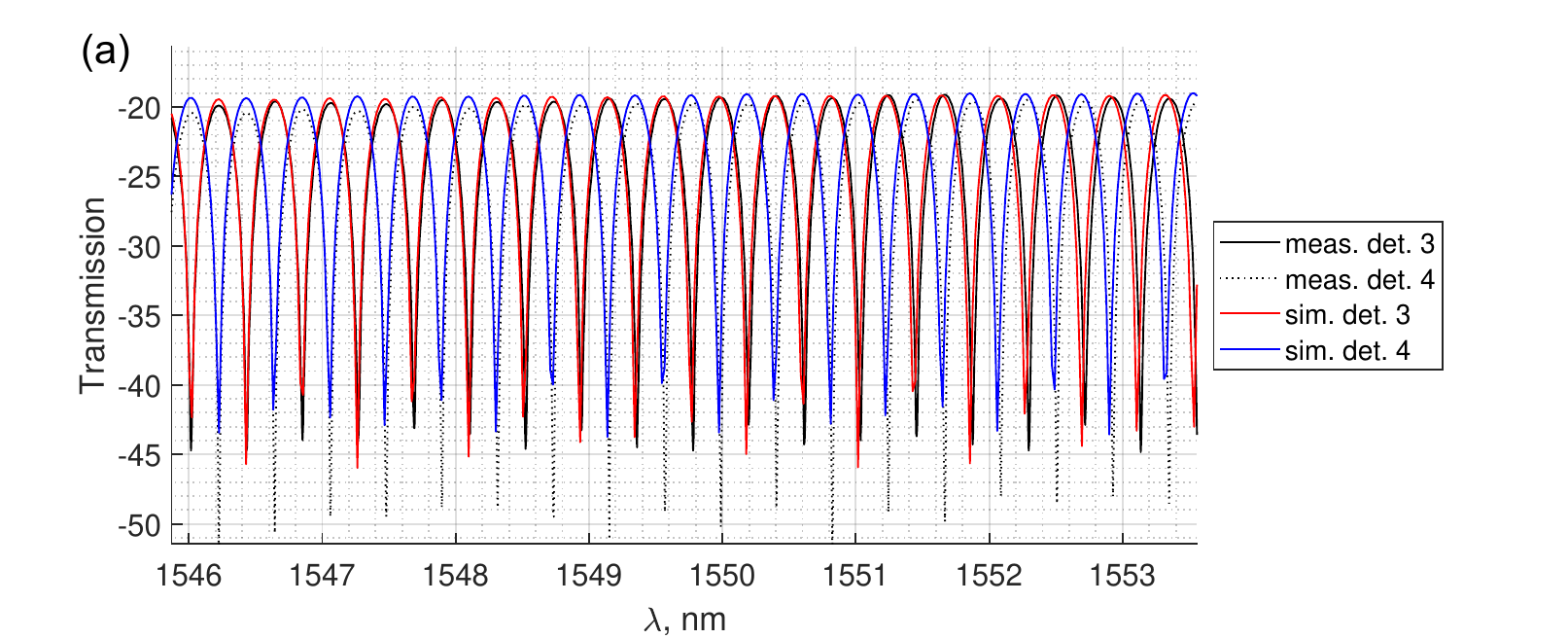}
    \includegraphics[width=0.49\linewidth]{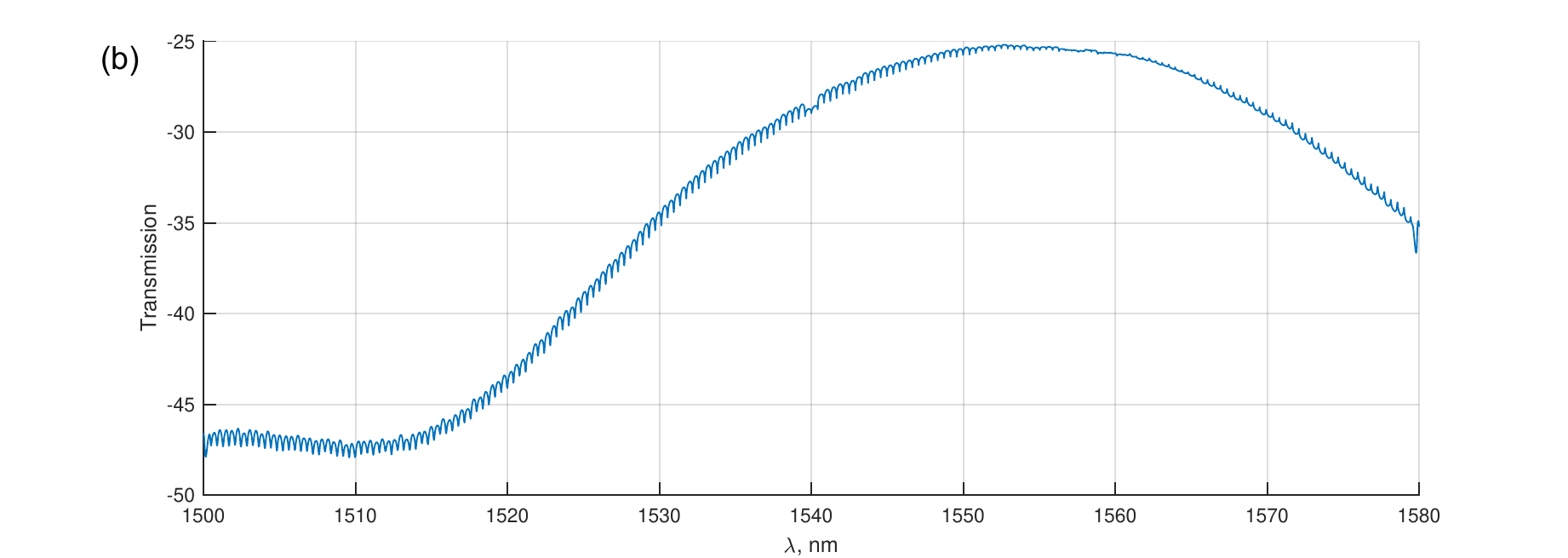}
    \includegraphics[width=0.49\linewidth]{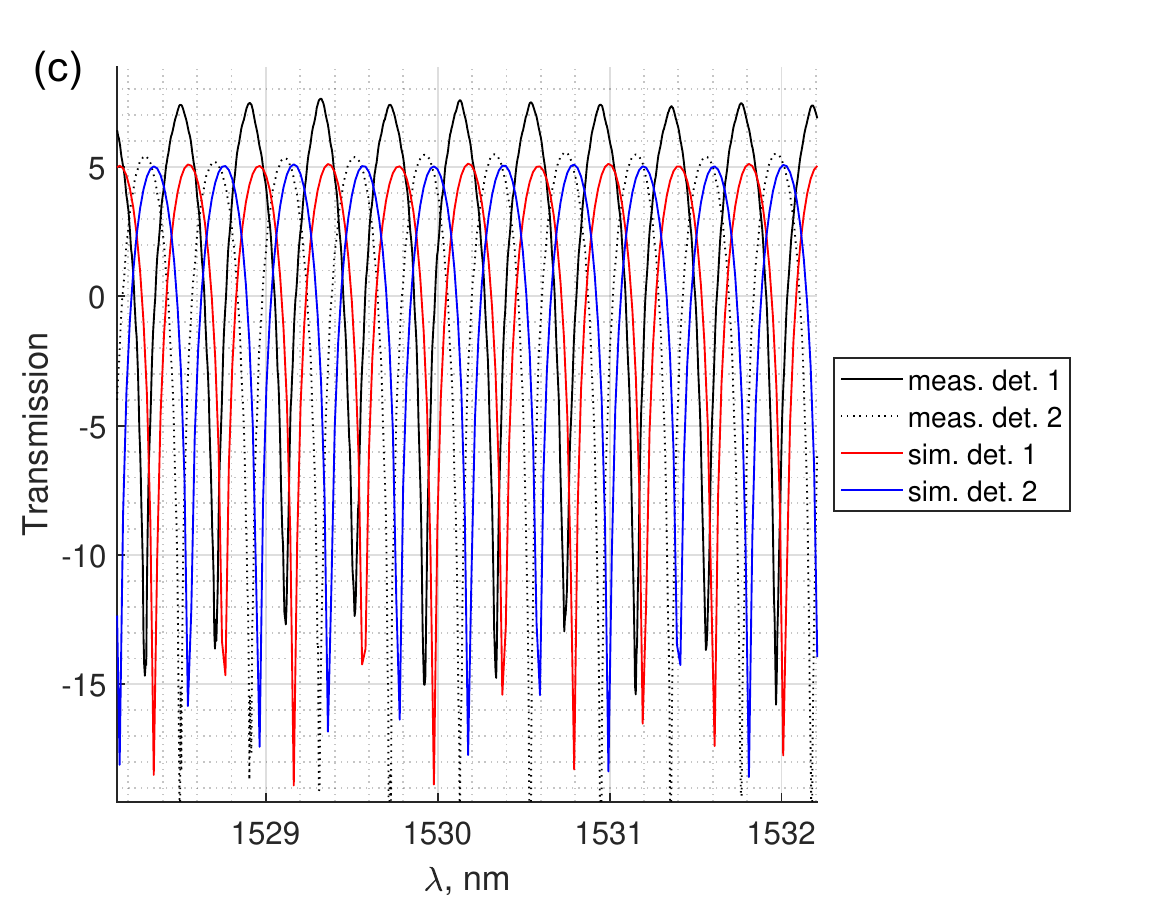} 
    \includegraphics[width=0.49\linewidth]{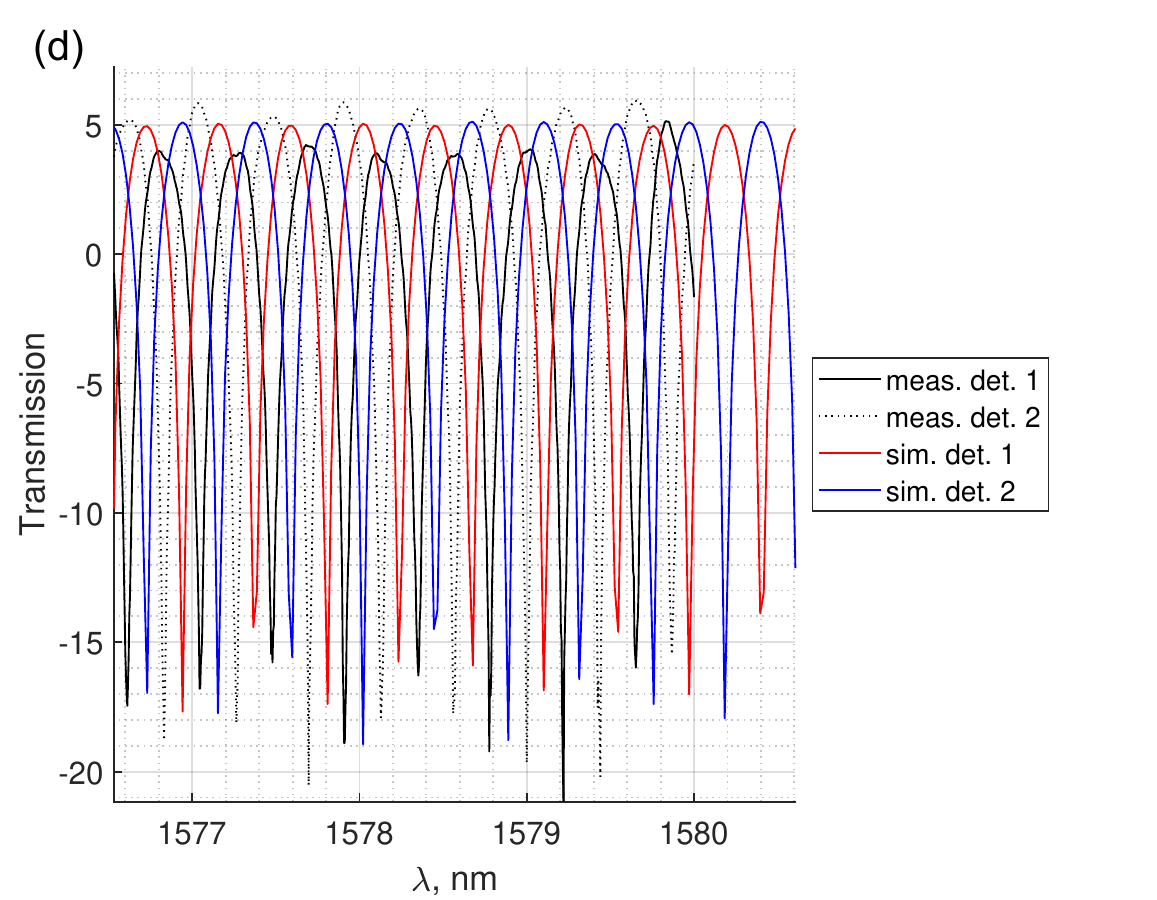}
    \caption{(a) The simulated and measured spectra from detectors 3 and 4 in narrow FSR MZI3\_1. There is an excellent agreement between the measured and simulated spectra in the wavelength range $\lambda=[1546,1554]$ \si{nm} illustrating that the desired free spectral range is achieved with negligible artifacts in the spectrum from process variation. Agreement between the spectra worsens and becomes opposite to the desired transmission at $\lambda=1530$ \si{nm} and $\lambda=1570$ \si{nm}, as shown in (c) and (d). Analyzing the spectra is difficult in this case without a de-embedding structure. Instead a moving mean was used, but does not represent the appopriate fitting parameters. Therefore, fitting is performed on the broadband MZIs later in this report.}
    \label{fig:9}
\end{figure}

\begin{figure}
    \centering
    \includegraphics[width=\linewidth]{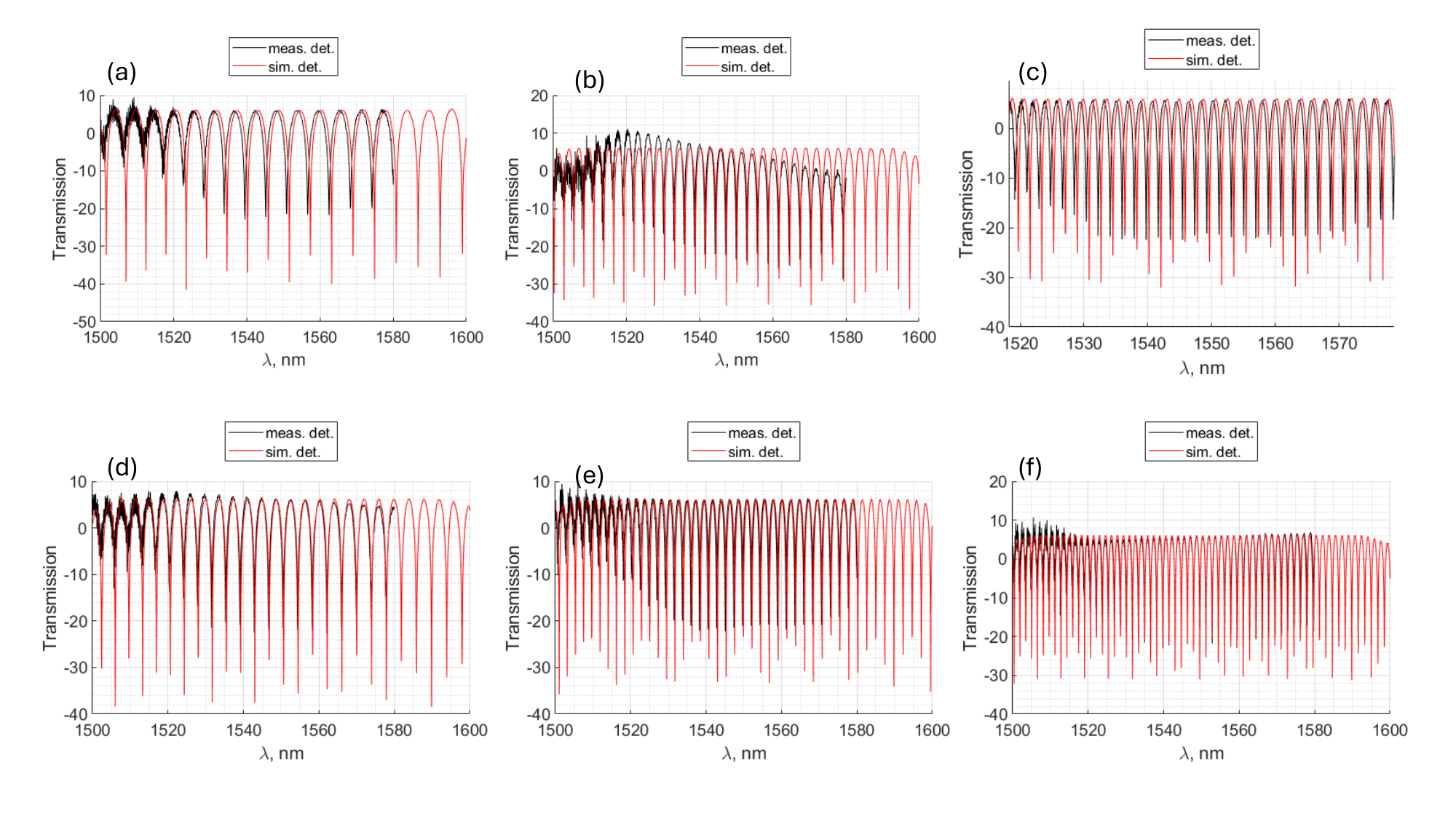}
    \caption{(a) MZI1\_2, (b) MZI2\_2, (c) MZI3\_2, (d) MZI4\_1, (e) MZI5\_1, and (f) MZI6\_1. Each shows excellent agreement with the simulated spectra from ANSYS Lumerical INTERCONNECT, but the spectra are shifted slightly for MZIs (a), (b), and (c). The spectral shift in (a), (b) and (c) may be associated with a change in $n_{\mathrm{eff}}$. In (d), (e) and (f), the measured and simulated spectra are an almost perfect fit. Baseline correction was used with reference to the YWGD structure in these figures, which does not appear relevant for each MZI. This suggests the process varied between the MZIs and the YWGD structure, such that a polynomial fit is more applicable for baseline subtraction.}
    \label{fig:11}
\end{figure}

There is excellent agreement between the FSR and path length mismatch in the wavelength range $\lambda=[1546,1554]$ \si{nm}. Fig. \ref{fig:9}(a) illustrates the excellent agreement between the simulated and measured transmission spectra after adding an insertion loss constant $b=-13.5710$ to every transmission data point. This design did not include a de-embedding structure, since the terminator element for the Y-branch did not satisfy the design rule checking which requested minimum feature size of 150 \si{nm} (while the end of the terminator was 60 \si{nm}). The DRC violation prohibited use of a de-embedding structure, although in a secondary layout (EBeam\_cael3.gds), the de-embedding structure did not satisfy baseline correction for all of the transmitted spectra (Fig. \ref{fig:11}). Instead, a good estimate can be made simply using either a polynomial fit or a moving mean average of the FSR window for simulated and  MZI spectra, to a similar effect as the de-embedding structure. This is demonstrated in Fig. \ref{fig:9}(b). In Figs. \ref{fig:9}(c) and \ref{fig:9}(d), the deviation between the spectra near $\lambda= 1530$ \si{nm} and $\lambda=1570$ \si{nm} is apparent, illustrating the narrow bandwidth of the selected FSR.

Fig. \ref{fig:12} shows that with baseline correction performed using a polynomial fit, as opposed to the de-embedding structure, the spectra from each MZI fitted using the MATLAB \texttt{findpeaks()} function has excellent correlation in the first attempt. In MZI1\_2, the wavelength shift was 0.71 \si{nm}, the center wavelength was $\lambda_0=1533.8$ \si{nm}, and the group refractive index estimate was $n_g=4.18301\pm0.00250$ as shown in Fig. \ref{fig:12}. Accurate group index is recovered using the analytical spectrum fitted with least squares regression.

